\documentclass[11pt,a4paper]{article}
\usepackage{jheppub}
\usepackage[utf8]{inputenc}
\usepackage{enumitem}
\usepackage{shuffle}
\usepackage{xcolor}
\allowdisplaybreaks

\usepackage{graphicx}
\usepackage[export]{adjustbox}

\def\be{\begin{equation}}
\def\ee{\end{equation}}
\def\bea{\begin{eqnarray}}
\def\eea{\end{eqnarray}}
\def\beal{\begin{equation}\begin{aligned}}
\def\eeal{\end{aligned}\end{equation}}
\def\nn{\nonumber}
\def\bra#1{\langle #1|}
\def\ket#1{|#1 \rangle}

\def\braket#1{\langle #1 \rangle}
\def\u#1{\underline{#1}}
\def\o#1{\overline{#1}}

\def\Res_#1{\operatorname*{Res}_{#1}}
\def\sgn{\operatorname*{sgn}}

\def\ie{i.~e. }
\def\eg{e.~g. }
\def\etc{etc}

\def\eqn#1{eq.~\eqref{#1}}
\def\eqns#1#2{eqs.~\eqref{#1} and~\eqref{#2}}

\def\fig#1{figure~{\ref{#1}}}

\def\Figs#1#2{Figures~{\ref{#1}} and~{\ref{#2}}}

\def\app#1{Appendix~{\ref{#1}}}
\def\rcite#1{\cite{#1}}
\def\rcites#1{\cite{#1}}

\newcommand{\aket}[1]{| #1 \rangle}
\newcommand{\abra}[1]{\langle #1 |}
\newcommand{\sket}[1]{| #1 ]}
\newcommand{\sbra}[1]{[ #1 |}
\newcommand{\abraket}[1]{\langle #1 \rangle}
\newcommand{\sbraket}[1]{[ #1 ]}
\newcommand{\abrasket}[3]{\langle #1 | #2 |#3 ]}
\newcommand{\sbraaket}[3]{[ #1 | #2 | #3 \rangle}

\newcommand{\sbrasket}[3]{[ #1 | #2 |#3 ]}

\title{All-multiplicity amplitudes with four massive quarks and identical-helicity gluons
}

\author[a]{Achilleas Lazopoulos,}
\author[b]{Alexander Ochirov}
\author[c]{and Canxin Shi}

\affiliation[a]{ETH Z\"urich, Institut f\"ur Theoretische Physik,
Wolfgang-Pauli-Str. 27, 8093 Z\"urich, Switzerland}
\affiliation[b]{Mathematical Institute, University of Oxford, \\
Andrew Wiles Building, Radcliffe Observatory Quarter,
Woodstock Road, Oxford OX2 6GG, U.K.}
\affiliation[c]{Institut f\"ur Physik und IRIS Adlershof,
Humboldt-Universit\"at zu Berlin, \\
Zum Gro{\ss}en Windkanal 2, 12489 Berlin, Germany}

\emailAdd{lazopoli@phys.ethz.ch}
\emailAdd{ochirov@maths.ox.ac.uk}
\emailAdd{canxin@physik.hu-berlin.de}

\preprint{SAGEX-21-21-E,
          HU-EP-21/47}

\abstract{We explore the on-shell recursion for tree-level scattering amplitudes with massive spinning particles. Based on the factorization structure encoded in the same way by two different recursion relations, we conjecture an all-multiplicity formula for two gauged massive particles of arbitrary spin and any number of identical-helicity gluons. Specializing to quantum chromodynamics (QCD), we solve the on-shell recursion relations in the presence of two pairs of massive quarks and an arbitrary number of identical-helicity gluons. We find closed-form expressions for the two distinct families of color-ordered four-quark amplitudes, in which all gluons comprise a single color-adjacent set. We compare the efficiency of the numerical evaluation of the two resulting analytic formulae against a numerical implementation of the off-shell Berends-Giele recursion. We find the formulae for both amplitude families to be faster for large multiplicities, while the simpler of the two is actually faster for any number of external legs. Our analytic results are provided in a computer-readable format as two ancillary files.
}

\begin{document}
\maketitle
\addtocontents{toc}{\protect\setcounter{tocdepth}{1}}

%%%%%%%%%%%%%%%%%%%%%%%%%%%%%%%%%%%%%%%%%%%%%%%%%%%%
\section{Introduction}
\label{sec:intro}
%%%%%%%%%%%%%%%%%%%%%%%%%%%%%%%%%%%%%%%%%%%%%%%%%%%%

Scattering amplitudes in gauge theory have been long known
to allow for drastic analytic simplifications,
in particular when expressed using the spinor-helicity formalism~\cite{Berends:1981rb,DeCausmaecker:1981bg,Gunion:1985vca,Kleiss:1985yh,Xu:1986xb,Gastmans:1990xh}.
The prime example is the tree amplitude for $n$ gluons,
of which two carry negative helicity~\cite{Parke:1986gb}:
\be
   A(1^+,\ldots,j^-,\ldots,l^-,\ldots,n^+)
    = \frac{i \braket{j\;\!l}^4}{\braket{1\;\!2} \braket{2\;\!3} \cdots
              \braket{n\!-\!1|n} \braket{n\;\!1}} ,
\label{MHV}
\ee
written here after color ordering (see \eg \rcite{Dixon:1996wi}).
This single-term amplitude,
also known as maximally helicity-violating (MHV),
is the epitome of simplicity hidden inside
a factorially growing avalanche of Feynman diagrams.
The all-plus and one-minus helicity configurations
are even simpler in that their amplitudes vanish.
On the other hand, configurations with three or more minus-helicity gluons
(NMHV, N$^2$MHV, etc.) involve multiple terms,
the number of which goes roughly like $n^k$ for an N$^k$MHV amplitude
\cite{Drummond:2008cr,Dixon:2010ik}.
Note that this is still significantly tamer
than the naively expected factorial growth.

The possibility for such on-shell analytic simplifications
can be attributed to
\begin{itemize}[leftmargin=\parindent]
\item gauge redundancies, which tie together different Feynman vertices;
\item fields with more indices than needed to describe
their on-shell spin degrees of freedom.
\end{itemize}
A natural way to sidestep such field-theoretic complications
and directly target streamlined amplitude expressions
is the purely on-shell method of Britto-Cachazo-Feng-Witten
(BCFW) recursion \cite{Britto:2004ap,Britto:2005fq},
which relates higher-point amplitudes to those at lower points.

Scattering amplitudes involving massive spinning particles
are typically more complicated than those in the purely massless sector.
However, the on-shell strategy outlined above
applies to them in its entirety
and can be implemented in the massive spinor-helicity formalism
of Arkani-Hamed, Huang and Huang~\cite{Arkani-Hamed:2017jhn}
(for earlier iterations see
\rcites{Kleiss:1986qc,Dittmaier:1998nn,Schwinn:2005pi,
Conde:2016vxs,Conde:2016izb}).

The first all-multiplicity results in this formalism
were obtained for tree-level QCD amplitudes with two massive quarks
by one of the current authors \cite{Ochirov:2018uyq}.
In this paper, we
\begin{enumerate}[leftmargin=\parindent]
\item generalize an $n$-point result of~\rcite{Ochirov:2018uyq}
to arbitrary spinning matter, see \eqn{MMggnAP};
\item find new all-multiplicity formulae~\eqref{eqAn1} and~\eqref{eqAn2}
for QCD amplitudes with four massive quarks and $(n-4)$
positive-helicity gluons.
\end{enumerate}
The latter results exhibit a pattern of increasing analytic complexity,
which is reminiscent of pattern seen in the massless N$^k$MHV amplitudes.
These observations lead us to
\begin{enumerate}[leftmargin=\parindent] \setcounter{enumi}{2}
\item explore if our analytic formulae are advantageous
to other methods in the context of purely numerical amplitude evaluation.
\end{enumerate}

Indeed, a widely used method for computing tree amplitudes
is the off-shell Berends-Giele (BG) recursion~\cite{Berends:1987me},
which is known to be very robust
and efficient in purely numerical calculations
\cite{Dinsdale:2006sq,Duhr:2006iq,Giele:2008bc,Ellis:2008qc,Lazopoulos:2008ex,Badger:2010nx,Badger:2012uz}.
In fact, in massless QCD it was shown~\cite{Badger:2012uz} to outperform
evaluation of closed formulae~\cite{Drummond:2008cr,Dixon:2010ik}
starting at the N$^2$MHV level of analytic complexity.
For QCD amplitudes with massive quarks,
we find that the evaluation of the simple two-quark formulae,
quoted from \rcite{Ochirov:2018uyq} in \eqns{QQggnAP}{QQggnOM},
is increasingly faster than the BG recursion,
as the total number of particles $n$ grows.
In the four-quark case, we observe a similar situation for one of the two
families of color-ordered amplitudes admitting closed-form expressions.
For the other such family, however,
our analytic results become numerically advantageous
only at multiplicities higher than $n=15$.
This suggests that the complexity of the considered,
relatively simple, massive external-particle configurations
is already approaching the edge of what it makes sense to tackle analytically,
assuming that improving the numerical efficiency
of tree-amplitude generation is what one is after.

In order to facilitate the reuse of our results
in analytical or numerical studies,
we provide the ancillary file \texttt{TreeAmpResults.txt}
with flexible Wolfram-friendly implementations
of the two-quark amplitude expressions of \rcite{Ochirov:2018uyq},
as well as the new four-quark formulae.
Our results involve composite auxiliary spinors,
a sample implementation of which can be found
in the second ancillary file \texttt{TreeAmpAuxSpinors.txt}.

%%%%%%%%%%%%%%%%%%%%%%%%%%%%%%%%%%%%%%%%%%%%%%%%%%%%
\section{From 3 to $\boldsymbol{n}$ points for general spin}
\label{sec:generalspin}
%%%%%%%%%%%%%%%%%%%%%%%%%%%%%%%%%%%%%%%%%%%%%%%%%%%%

In this section we describe the on-shell approach to
tree-level amplitudes with massive particles
and demonstrate its effectiveness by deriving
an all-multiplicity all-spin formula for gauged spinning matter.

In the spinor-helicity formalism~\cite{Arkani-Hamed:2017jhn},
the basic building blocks for scattering amplitudes
are massive $2 \times 2$ spinors denoted by angle or square bras and kets:\footnote{For a comprehensive exposition of the formalism,
we refer the reader to \rcite{Arkani-Hamed:2017jhn}. Our conventions
are consistent with the latest arXiv version of \rcite{Ochirov:2018uyq}.
For completeness, in \app{app:Parametrizations}
we provide explicit spinor parametrizations in terms of momentum components,
which can be used for numerical evaluation.}
\be
   \ket{p^a}_{\;\!\!\alpha}\;\![p_a|_{\dot{\alpha}}
    = \epsilon_{ab} \ket{p^a}_{\;\!\!\alpha}\;\![p^b|_{\dot{\alpha}}
    = p_{\alpha\dot{\alpha}}
    = p_\mu \sigma^\mu_{\alpha\dot{\alpha}} , \qquad \quad
   p^2\! = \det\{p_{\alpha\dot{\alpha}}\} = m^2 \neq 0 .
\ee
They carry the ${\rm SU}(2)$ little-group indices~$a$ --- in addition to
the ${\rm SL}(2,\mathbb{C})$ Weyl indices~$\alpha$ and~$\dot{\alpha}$
representing the Lorentz group,
which are equally relevant for massless 2-spinors
\be
   \ket{k}_{\alpha} [k|_{\dot{\alpha}}
    = k_{\alpha \dot{\alpha}}
    = k_\mu \sigma^\mu_{\alpha\dot{\alpha}} , \qquad \quad
   k^2\!= \det\{k_{\alpha\dot{\alpha}}\} = 0 .
\ee
The massless little group is ${\rm U}(1)$, so it requires
no additional indices besides the complex structure of the spinors.
The spins of the massless particles
are therefore represented by helicity weights:
by convention we assign helicity $-1/2$ to $\ket{k}$ and $+1/2$ to $|k]$.
Moreover, each spin-$s$ massive particle is represented
by $2s$ symmetrized ${\rm SU}(2)$ indices.
All ${\rm SL}(2,\mathbb{C})$ indices are always contracted
in an on-shell amplitude.

%%%%%%%%%%%%%%%%%%%%%%%%%%%%%%%%%%%%%%%%%%%%%%%%%%%%
\subsection{General gauged-matter amplitude}
\label{sec:generalspinAP}
%%%%%%%%%%%%%%%%%%%%%%%%%%%%%%%%%%%%%%%%%%%%%%%%%%%%

Let us use the formalism outlined above to write the scattering amplitude
for a massive quark-antiquark pair and $(n-2)$ positive-helicity gluons
\cite{Ochirov:2018uyq}:
\be\!\!
   A(\u{1}^a,2^+\!,3^+\!,\dots,(n-1)^+\!,\o{n}^b)
    = \frac{ i\:\!m \braket{1^a n^b}
             [2| \prod_{j=2}^{n-3}\!
                 \big\{\!\!\not{\!\!P}_{12 \dots j}\!\not{\!p}_{j+1}
                      + (s_{12 \dots j}-m^2) \big\} |n\!-\!1] }
           {\!\!\!\prod_{j=2}^{n-2} \braket{j|j\!+\!1} (s_{12 \dots j}-m^2)\;} ,\!\!
\label{QQggnAP}
\ee
where we label spinors by particle numbers instead of momenta,
\ie $\ket{p_j}_\alpha = \ket{j}_\alpha$.
By convention, we consider all momenta outgoing
and denote momentum sums as $P_{12 \dots j} = p_1 + p_2 + \ldots + p_j$
and their Lorentz squares as $s_{12 \dots j}$.
Slashed matrices $\not{\!\!P}$ mean either $P_{\alpha\dot{\alpha}}$ or
$ P^{\dot{\alpha}\alpha} = \epsilon^{\alpha\beta}
  \epsilon^{\dot{\alpha}\dot{\beta}} P_{\beta\dot{\beta}} $
depending on the spinors surrounding them.
Hence the numerator in \eqn{QQggnAP} contains $(n-4)$ factors of
$\big\{(P_{12 \dots j})^{\dot{\alpha}\gamma}(p_{j+1})_{\gamma\dot{\beta}}
+(s_{12 \dots j}-m^2) \delta^{\dot{\alpha}}_{\dot{\beta}}\big\}$.
Their order of matrix multiplication is such that
$j$ increases from left to right.

Similarly to the MHV amplitude~\eqref{MHV},
the two-quark amplitude above consists of a single term
for any number of plus-helicity gluons.
Moreover, it depends on the quark spin labels~$a$ and~$b$
only through the single spinor product
$\braket{1^a n^b} = \epsilon^{\beta\alpha} \ket{1^a}_\alpha\ket{n^b}_\beta$.
In fact, the amplitude with two massive scalars instead of quarks,
derived in \rcites{Forde:2005ue,Ferrario:2006np,Schwinn:2007ee},
can be obtained by mere replacement $\braket{1^a n^b} \to m$.
We can therefore make a good educated guess for the amplitude
with $(n-2)$ plus-helicity gluons and two massive spin-$s$ particles:\footnote{The
general-spin formula~\eqref{MMggnAP} first appeared in an explicit form
in the lectures~\cite{Ochirov:2020lect}
and implicitly already in \rcite{Aoude:2020onz} in the context of
heavy-particle effective theory.
Apart from matching the $s=0,1/2$ results
of \rcites{Ferrario:2006np,Schwinn:2007ee,Ochirov:2018uyq},
it has also been recently derived for $s=1$ in \rcite{Ballav:2021ahg}.
}
\beal\!\!\!
   A(\u{1}^{\{a_1,\dots,a_{2s}\}}\!,2^+\!,3^+\!,&\dots,(n-1)^+\!,
     \o{n}^{\{b_1,\dots,b_{2s}\}}) \\ &
    = (-1)^{\lfloor s \rfloor}
      \frac{ i \braket{1^a n^b}^{\odot 2s}
             [2| \prod_{j=2}^{n-3}\!
                 \big\{\!\!\not{\!\!P}_{12 \dots j}\!\not{\!p}_{j+1}
                      + (s_{12 \dots j}-m^2) \big\} |n\!-\!1] }
           { m^{2s-2} \prod_{j=2}^{n-2} \braket{j|j\!+\!1}
             (s_{12 \dots j}-m^2) } .\!
\label{MMggnAP}
\eeal
Here and below $\braket{1^a n^b}^{\odot 2s}$ is a shorthand for
the tensor product symmetrized in the (separate) ${\rm SU}(2)$ index sets
$\{a_i\}$ and $\{b_i\}$
\begin{equation}
   \braket{1^a n^b}^{\odot 2s}
    = \braket{1^{(a_1} n^{(b_1}} \braket{1^{a_2} n^{b_2}} \cdots
      \braket{1^{a_{2s})} n^{b_{2s})}} .
\end{equation}
The sign prefactors $(-1)^{\lfloor s \rfloor}$ are included in \eqn{MMggnAP}
for the sake of consistency with \rcite{Johansson:2019dnu}.
The massive particles are understood to transform
in some representation of an arbitrary gauge group.
As long as we are allowed to pick the adjoint representation of ${\rm SU}(N_c)$
or ${\rm U}(N_c)$, we have a well-defined notion of color ordering.
Then the permutations of \eqn{MMggnAP} may be color-dressed
as required by any other group and representation
\cite{DelDuca:1999rs}.
For instance, in the case of ${\rm U}(1)$
we have the following amplitude with $(n-2)$ photons:
\beal
   {\cal A}(1^{\{a\}}\!,2,3,\dots,n-1,n^{\{b\}})
    = \big(\sqrt{2}Q\big)^{n-2}\!\!\!\!\!
      \sum_{\sigma \in S_{n-2}(\{2,3,\dots,n-1\})}\!\!\!\!\!
      A(1^{\{a\}}\!,\sigma,n^{\{b\}}) ,
\label{QCD2QED}
\eeal
where $Q$ is the charge of the two charged matter particles.
The spin-1/2 case simply corresponds to quantum electrodynamics.
Perhaps more interestingly,
the spin-1 version of \eqns{MMggnAP}{QCD2QED}
gives a closed-form all-multiplicity expression for the amplitude
involving two $W$ bosons emitting $(n-2)$ plus-helicity photons,
as follows from the recursive arguments below.
Furthermore, the gravitational analogue of the above amplitudes,
in which the massless gauge bosons are replaced by positive-helicity gravitons,
can be obtained via the Kawai-Lewellen-Tye (KLT) double copy~
\cite{Kawai:1985xq,Bern:1998sv}.
Indeed, in the case of two matter particles,
the massive-matter extension of the double copy
\cite{Johansson:2015oia,delaCruz:2016wbr,Brown:2018wss,Johansson:2019dnu}
is no more complicated than in the purely massless case
and therefore has a simple closed form \cite{Bern:1998sv}.

The spin-$1/2$ formula~\eqref{QQggnAP} was proven \cite{Ochirov:2018uyq}
recursively in QCD with the four-point amplitude as the starting point.
The induction step was set up using a massless BCFW shift
\be
   |\widehat{n\!-\!2}] = |n\!-\!2] - z |n\!-\!1] , \qquad \quad
   \ket{\widehat{n\!-\!1}} = \ket{n\!-\!1} + z \ket{n\!-\!2} ,
\label{BCFWshiftn2n1}
\ee
which by construction preserves the on-shell conditions
$\hat{p}_{n-2}^2 = \hat{p}_{n-1}^2 = 0$ and momentum conservation
$\hat{p}_{n-2} + \hat{p}_{n-1} = p_{n-2} + p_{n-1}$.
Proving the spin-$s$ formula~\eqref{MMggnAP}
can be approached in the same way, starting from
the corresponding four-point result of \rcite{Johansson:2019dnu}
and assuming the so-called ``minimal-coupling'' three-point amplitude
\cite{Arkani-Hamed:2017jhn} for the induction step:
\be
   A(\u{1}^{\{a\}}\!,2^+,\o{3}^{\{b\}})
    = (-1)^{\lfloor s \rfloor+1}
      \frac{i \braket{1^a 3^b}^{\odot 2s} [2|1\ket{q} }
           { m^{2s} \braket{2\;\!q} } ,
\label{MMgPlus}
\ee
where $\ket{q}$ is an arbitrary reference spinor.
We can then effortlessly generalize the spin-$1/2$ derivation
to the spin-$s$ case.
Instead of showing here essentially the same calculations
as in \rcite{Ochirov:2018uyq},
let us rather discuss an important remaining issue regarding higher spins,
which will motivate the need for an independent rederivation of \eqn{MMggnAP}.

Recall that BCFW recursion \cite{Britto:2004ap,Britto:2005fq}
treats a complex-shifted tree amplitude as a rational function of $z$
and relates the residues of finite poles in $z$ to lower-point amplitudes.
Now the spin-$1/2$ derivation of \rcite{Ochirov:2018uyq} relied
on the vanishing behavior of the $n$-point amplitude
at $z \to \infty$, which is guaranteed for QCD
with massive or massless quarks
\cite{Badger:2005jv,Britto:2012qi}.
It is also known to be true for gauge theory coupled to scalar matter
\cite{Badger:2005zh},
as well as for a spontaneously broken gauge theory
\cite{Badger:2005jv}.
In the latter case, it follows from the irrelevance
of the lower-energy symmetry-breaking effects
for the large-momentum behavior of the tree amplitude.
so the classical arguments in pure gauge theory
(see \eg \rcite{ArkaniHamed:2008yf}) still hold for suitably chosen shifts.
In the case where the unbroken part of the gauge symmetry is ${\rm U}(1)$,
the sum over the orderings in \eqn{QCD2QED} must result
in an amplitude for a $W$-boson pair emitting $(n-2)$ photons.

The problematic point is that tame boundary behavior
is not a given for higher spins.
In fact, for generic higher-spin amplitudes it must fail,
because a naive BCFW derivation of the Compton scattering amplitude
${\cal A}(1^{\{a\}}\!,2^+,3^-,4^{\{b\}})$ with opposite helicities
is already known \cite{Johansson:2019dnu} to produce an expression
with unphysical pole structure unless $s \leq 1$.
One could then proceed to search for
additional amplitude contributions which vanish on the physical poles
but subtract out the unphysical pole from the naive BCFW result
\cite{Arkani-Hamed:2017jhn,Chung:2018kqs,Falkowski:2020aso,Chiodaroli:2021eug}.

In the case of the all-plus formula~\eqref{MMggnAP}, however,
we have reasons to suspect that it constitutes
a valid spin-$s$ amplitude by itself:
\begin{itemize}

\item First of all, the BCFW-shifted form of the all-plus amplitude
behaves as ${\cal O}(1/z)$ for $z \to \infty$.
Indeed, the denominator contains a factor of
$\hat{s}_{12\dots(n-2)}-m^2 = {\cal O}(z)$,
whereas the numerator dependence on $z$ is canceled in
$\hat{\not{\!p}}_{n-2} |n\!-\!1] = \ket{n-2} [n\!-\!2|n\!-\!1]$.
Therefore, the proposed formula is consistent
with the vanishing boundary behavior.

\item Moreover, the only lower-point input needed to build up
the all-plus formula is the three-point amplitude~\eqref{MMgPlus}
with the positive-helicity gluon.
In other words, the amplitude~\eqref{MMggnAP}
is sensitive exclusively to how the matter is coupled
to the self-dual sector of gauge theory.
Therefore, if one imagines an effective field theory,
from which this formula follows,
it may seem plausible that
the higher-dimension effective operators in this theory
(\eg of the schematic form $\bar{\Phi} F^{n-2} \Phi/m^{2n-6}$)
would behave in a particular way
when evaluated on self-dual gauge fields $F=i\,{}^*\!F$,
which would lead to a vanishing boundary behavior of the all-plus amplitude.
Constructing a field-theoretic argument along these lines would be
very interesting but would go outside of the on-shell scope of this paper.

\item Finally, we can apply a different BCFW shift
and see that it is consistent with the same formula~\eqref{MMggnAP}.
Let us construct such an argument below.

\end{itemize}

%%%%%%%%%%%%%%%%%%%%%%%%%%%%%%%%%%%%%%%%%%%%%%%%%%%%
\subsection{Massive-massless BCFW shift}
\label{sec:massiveBCFW}
%%%%%%%%%%%%%%%%%%%%%%%%%%%%%%%%%%%%%%%%%%%%%%%%%%%%

Although the BCFW recursion was originally established
via shifting two massless momenta \cite{Britto:2005fq},
it was soon extended to shifts involving massive momenta
\cite{Badger:2005zh,Schwinn:2007ee}.
At first the corresponding massive spinors used to be defined using
the massless spinor-helicity formalism with the help
of an additional massless reference momentum
\cite{Kleiss:1986qc,Dittmaier:1998nn,Schwinn:2005pi}.
The new massive spinor-helicity formalism \cite{Arkani-Hamed:2017jhn}
allows for a more elegant formulation of such shifts,
as explored in \rcites{Herderschee:2019dmc,Aoude:2019tzn,
Franken:2019wqr,Falkowski:2020aso,Ballav:2020ese}.
In the rest of this paper,
we will be picking one massive particle~$j$ and one massless particle~$k$
and shift their on-shell spinors as follows \cite{Badger:2005zh,Aoude:2019tzn}:
\be
   |\hat{j}^a] =|j^a] - \frac{z}{m} |k] [k\;\!j^a] , \qquad \quad
   \ket{\hat{k}} = \ket{k} - \frac{z}{m}\!\not{\!p}_j|k] ,
\label{MassiveMasslessShift}
\ee
where the mass~$m$ of particle~$j$ is used to ensure
that $z$ is dimensionless.
In terms of momenta, this $[j^a,k\rangle$ shift translates to
\be
   \hat{p}_j = p_j - z r , \qquad \quad
   \hat{p}_k = p_k + z r , \qquad \quad
   r^\mu = \frac{1}{2m}[k|\!\!\not{\!p}_j\sigma^\mu|k] .
\label{MassiveMasslessShiftMomenta}
\ee
One can immediately see that the momentum conservation and
on-shell conditions are preserved, $\hat{p}_j^2=m^2$ and $\hat{p}_k^2=0$.

%%%%%%%%%%%%%%%%%%%%%%%%%%%%%%%%%%%%%%%%%%%%%%%%%%%%
\paragraph{Alternative derivation of spin-$\boldsymbol{s}$ amplitude.}
Let us immediately employ this shift to construct an inductive argument
for the spin-$s$ amplitude~\eqref{MMggnAP}.
First of all, note that if we choose to apply it to the last two particles,
the factor of $\braket{1^a n^b}^{2s}$ stays unaffected
by the $[n^b,n-1\rangle$ shift, as does the rest of the formula
except for the denominator factor $\braket{n\!-\!2|\widehat{n\!-\!1}}$.
Therefore, the conjectured expression behaves as ${\cal O}(1/z)$
for $z \to \infty$,
similarly to the situation with the massless shift~\eqref{BCFWshiftn2n1}.
This lets us continue to assume a vanishing boundary behavior
in our construction of the induction step.

Considering the $[n^b,n-1\rangle$ shift more closely,
we see that every color-ordered factorization channel
which separates the shifted particles necessarily involves
a purely gluonic amplitude with at most one negative helicity.
The only case where such an amplitude does not vanish
is when it is trivalent and evaluated on complex kinematics.
Therefore, the only pole with a nonzero residue is produced
in the $\hat{s}_{(n-2)(n-1)}$-channel:
%\vspace{-5pt}
\begin{align}
   \includegraphics[valign=c,scale=0.98]{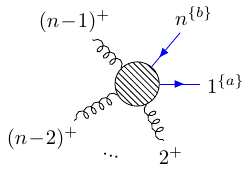} =
   \includegraphics[valign=c,scale=0.98]{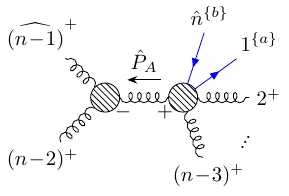} .
\label{MMggnAPrecursion}
\end{align}
To compute the left-hand side of this recursion,
we only need the three-gluon $\overline{\text{MHV}}$,
given for completeness in \app{app:BuildingBlocks}, as well as
assume that \eqn{MMggnAP} holds for the $(n-1)$-point spin-$s$ amplitude,
which constitutes our induction hypothesis. We find
%\vspace{-10pt}
\be\!
   \includegraphics[valign=c,scale=0.98]{./Graphs/graph26.pdf}\!
   \begin{aligned} \phantom{\Bigg|} \\
    = (-1)^{\lfloor s \rfloor}
      \frac{ \braket{1^a n^b}^{\odot 2s}
             [2| \prod_{j=2}^{n-4}\!
                 \big\{\!\!\not{\!\!P}_{12 \dots j}\!\not{\!p}_{j+1}
                      + (s_{12 \dots j}-m^2) \big\} |\hat{P}_A] }
           { m^{2s-2} \braket{2\;\!3} \cdots \braket{n\!-\!3|\hat{P}_A}
             \prod_{j=2}^{n-3} (s_{12 \dots j}-m^2) } & \\
      \phantom{_{\Big|}} \times\,
      \frac{-i}{s_{(n-2)(n-1)}} \times
      \frac{[n\!-\!2|n\!-\!1]^3}{[\hat{P}_A|n\!-\!2][n\!-\!1|\hat{P}_A]} & .
   \end{aligned}
\label{MMggnAPrecursion2}
\ee
%\vspace{-3pt}
On this pole, the shift variable takes the value
\be
   z = \frac{m \braket{n\!-\!2|n\!-\!1}\!}{\bra{n\!-\!2}n|n\!-\!1]},
   \qquad
   \ket{\hat{P}_A}[\hat{P}_A| = \ket{n\!-\!2}
      \frac{ [n\!-\!1|n|\!\not{\!\!P}_{(n-1)(n-2)} }
           { \bra{n\!-\!2}n|n\!-\!1] } .
\ee
(Here and below we use the notation like
$\bra{n\!-\!2}n|n\!-\!1]=\bra{n\!-\!2}\!\!\not{\!p}_n|n\!-\!1]$
interchangeably.)
Hence we may identify the denominator factor
$\braket{n\!-\!3|\hat{P}_A}=\braket{n\!-\!3|n\!-\!2}$, and furthermore
\be
   [\hat{P}_A|n\!-\!2] = \frac{(s_{(n-1)n}-m^2)}{\bra{n\!-\!2}n|n\!-\!1]}
      [n\!-\!1|n\!-\!2] , \qquad \quad
   [n\!-\!1|\hat{P}_A] = [n\!-\!1|n\!-\!2] .
\ee
Finally, rewriting the right-handed spinor
for the complex intermediate momentum $\hat{P}_A$ as
\be
   |\hat{P}_A]
    = \frac{ \big\{\!\!\not{\!\!P}_{12 \dots (n-3)}\!\not{\!p}_{n-2}
                      + (s_{12 \dots (n-3)}-m^2) \big\} |n\!-\!1] }
           { \bra{n\!-\!2}n|n\!-\!1] } ,
\ee
we conclude that the $n$-point residue~\eqref{MMggnAPrecursion2}
coincides with the conjectured expression~\eqref{MMggnAP}.

We now have two different derivations of the conjectured
all-multiplicity all-spin formula for gauged spinning matter,
which is governed by a single gauge coupling and
exhibits a vanishing boundary behavior for the used BCFW shifts.
All the poles in this amplitude are physical, with correct factorization channels
all the way down to the basic three-point amplitudes~\eqref{MMgPlus}.
In these respects, it is unique and must belong to an interesting family of
effective field theories of massive higher-spin matter
coupled to a gauge field.
For lower spins $s \leq 1$, the validity of \eqn{MMggnAP} is guaranteed
by the previously established results for scalars~\cite{Ferrario:2006np}
and quarks~\cite{Schwinn:2007ee,Ochirov:2018uyq},
as well as by the boundary-behavior arguments
of \rcites{ArkaniHamed:2008yf,Ballav:2020ese} for vector particles
in the context of a spontaneously broken gauge theory.

%%%%%%%%%%%%%%%%%%%%%%%%%%%%%%%%%%%%%%%%%%%%%%%%%%%%
\section{QCD amplitudes with two quark pairs}
\label{sec:amplitudes}
%%%%%%%%%%%%%%%%%%%%%%%%%%%%%%%%%%%%%%%%%%%%%%%%%%%%

In this section we proceed to the main computations of this paper:
four-quark amplitudes with any number of gluons of positive helicity.
We consider the more general case of two distinctly flavored quark pairs.
The identical-flavor amplitude may always be obtained
by setting the two masses to be equal and subtracting two relabelings:
\be
   {\cal A}({\color{blue} \u{1}},{\color{blue} \o{2}},
            {\color{blue} \u{3}},{\color{blue} \o{4}},5,\dots,n)
 = {\cal A}({\color{red} \u{1}},{\color{red} \o{2}},
            {\color{blue} \u{3}},{\color{blue} \o{4}},5,\dots,n)
 - {\cal A}({\color{red} \u{1}},{\color{blue} \o{2}},
            {\color{blue} \u{3}},{\color{red} \o{4}},5,\dots,n) ,
\ee
where we have color-coded the two quark flavors by red and blue.

Moreover, we concentrate on the color-ordered versions
of the tree amplitudes in question.
There are multiple possible color decompositions
\cite{Johansson:2015oia,Ochirov:2019mtf}
which can be used to assemble the full color-dressed amplitude
from different color orderings.
Let us first specialize to orderings of the form
$A({\color{red} \u{1}},{\color{red} \o{2}},
 \{{\color{blue} \u{3}},{\color{blue} \o{4}}\}\shuffle\sigma(5,\dots,n))$
where $\sigma \in S_{n-4}$ is an arbitrary permutation
and $\shuffle$ is the shuffle product.
In this setting, $\{{\color{blue} 3},{\color{blue} 4}\}\shuffle\sigma(5,\dots,n)$
amounts to all permutations of $\{{\color{blue} 3},{\color{blue} 4},5,\dots,n\}$,
in which the relative ordering of the quark labels~3 and~4 stays fixed.
These orderings constitute Melia's amplitude basis
\cite{Melia:2013bta,Melia:2013epa}
with respect to the Kleiss-Kuijf (KK) relations \cite{Kleiss:1988ne}.
This basis and the corresponding color decomposition
\cite{Johansson:2015oia,Melia:2015ika}
allows us to fix the relative position of the four quarks
and avoid gluon insertions between quarks~1 and~2.
Furthermore, the color-kinematic information additionally provided by
(the mass- and flavor-extended version of)
the Bern-Carrasco-Johansson relations
\cite{Bern:2008qj,Johansson:2015oia,delaCruz:2015dpa}
lets us consider only orderings of the form
$A({\color{red} \u{1}},{\color{red} \o{2}},{\color{blue} \u{3}},
   \sigma({\color{blue} \o{4}},5,\dots,n))$.

Color-ordered amplitudes contain poles
only in subsets of consecutively ordered momenta.
This is very helpful for solving on-shell recursions.
Indeed, the BCFW relations receive contributions from $z$-dependent poles,
with the $z$-shifted particles on the different sides
of each factorization channel.
Consequently, if two adjacent particles are shifted
in an $n$-point ordered amplitude, there are only
$n-3$ factorization channels to consider.

Note that some ordered amplitudes involving massive and massless particles,
such as
$A({\color{red} \u{1}},{\color{red} \o{2}},{\color{blue} \u{3}},{\color{blue} \o{4}},5)$,
lack a pair of adjacent massless particles,
so the purely massless shift of the type~\eqref{BCFWshiftn2n1}
cannot be applied to them.
This is why we find the massive-massless shift \eqref{MassiveMasslessShift}
especially useful for computing massive four-quark amplitudes.
Since we only consider gluons of positive helicity,
we shift the left-handed massless spinors, \eg
\be
   |\hat{4}^d] =\sket{4^d} - \frac{z}{m}\sket{5} [5\;\!4^d] , \qquad \quad
   \aket{\hat{5}} = \aket{5}- \frac{z}{m}|4\sket{5} .
\label{Shift45}
\ee
Note that in the limit where the quark 4 is taken to be massless
its ${\rm SU}(2)$ labels $d=1,2$ may be
associated to the positive and negative helicities, respectively.
Therefore, the above $[4^d,5\rangle$ shift
actually corresponds to two massless BCFW shifts:
$[4^+,5^+\rangle$ and $[4^-,5^+\rangle$.
As regards the $[4^-,5^+\rangle$ shift, in QCD it is easy to see
\cite{Luo:2005rx} that every Feynman diagram vanishes at infinity
no worse than the gluon's polarization vector
$\hat{\varepsilon}_5^+ = {\cal O}(1/z)$.
Showing the vanishing boundary behavior under
the massless $[4^+,5^+\rangle$ shift is already more involved
as the naive power counting estimates the amplitude to be ${\cal O}(1)$
\cite{Badger:2005zh,Schwinn:2007ee}.
For completeness, in \app{app:BoundaryBehavior}
we provide an argument based on the QCD Feynman rules
showing that for the general massive-massless shift $[j^a,k\rangle$
the leading ${\cal O}(1)$ terms vanish,
thus ensuring the applicability of on-shell recursion.

%%%%%%%%%%%%%%%%%%%%%%%%%%%%%%%%%%%%%%%%%%%%%%%%%%%%
\subsection{Gluon insertions between like-flavored quarks}
\label{sec:AmpType1}
%%%%%%%%%%%%%%%%%%%%%%%%%%%%%%%%%%%%%%%%%%%%%%%%%%%%

We have found the simplest family of color orderings
to be the amplitudes of the form
$A(\u{1}^a,\o{2}^b,\u{3}^c,5^+,\dots,n^+,\o{4}^d)$,
where quarks 1 and 2 carry mass $M$,
while quarks 3 and 4 have another flavor and mass $m$.
For concreteness, we set $1$ and $3$ to be outgoing quarks
and $2$ and $4$ to be antiquarks.
This can be easily tweaked, since reversing a fermionic line
in the color-ordered amplitude (without changing the ordering)
simply results in an overall minus sign,
as shown in \app{app:FermionReversal}.
We will be using this way of flipping quark lines to our convenience
in the rest of the paper.
Moreover, we will henceforth underscore the quarks and bar the antiquarks
only when their choice is not otherwise obvious,
which will unclutter the amplitude notation.

%%%%%%%%%%%%%%%%%%%%%%%%%%%%%%%%%%%%%%%%%%%%%%%%%%%%
\paragraph{5-point amplitude.}
First, let us compute the five-particle amplitude $A(1^a,2^b,3^c,5^+\!,4^d)$
using the aforementioned $[4^d,5\rangle$ shift in \eqn{Shift45}.
Due to flavor conservation,
this BCFW shift produces a single gauge-invariant contribution:
%\vspace{-5pt}
\be\!\!\!\!\!\!\!
   \includegraphics[valign=c,scale=0.9]{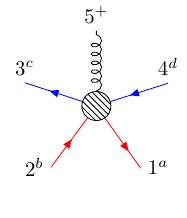}\!\!
    =\!\!\!\includegraphics[valign=c,scale=0.9]{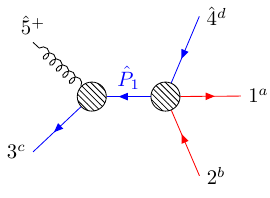}
    = A(3^c\!,\hat{5}^+\!,-\hat{P}_1^e) \frac{i}{s_{35}-m^2}
   A(1^a\!,2^b\!,\hat{P}_{1,e},\hat{4}^d) .\!
\label{grA12354}
\ee
The basic building blocks here are the lower-point amplitudes
\eqref{eq3pt} and \eqref{eq4Fermion} given in \app{app:BuildingBlocks}.
They combine into
\beal\!
  A(1^a\!,2^b\!,3^c\!,5^+\!,4^d)
   = \frac{ i \sbraaket{5}{3}{q} }
          { m s_{12} (s_{35}\!-\!m^2) \braket{\hat{5}\;\!q} }
     \braket{3^c \hat{P}_1^e}
     \Big( \braket{1^a \hat{P}_{1,e}} \sbraket{2^b \hat{4}^d}
         + \braket{1^a 4^d} \sbraket{2^b \hat{P}_{1,e}}
         + (1 \leftrightarrow 2) \Big) & \\
   = \frac{-i \sbraaket{5}{3}{q} }
          { s_{12} (s_{35}\!-\!m^2) \braket{\hat{5}\;\!q} }
     \bigg(\:\!\!\braket{1^a 3^c} \sbraket{2^b \hat{4}^d}
          + \braket{1^a 4^d} \sbraket{2^b 3^c}
          + \frac{1}{m} \braket{1^a 4^d} \sbraket{2^b 5}
            \braket{\hat{5}\;\!3^c}
          + (1 \leftrightarrow 2)\:\!\!\bigg) & ,
\label{grA12354compute}
\eeal
where we have eliminated $\hat{P}_1$ by using
$\ket{\hat{P}_{1,e}} \bra{\hat{P}_1^e} = m$ and
$|\hat{P}_{1,e}] \bra{\hat{P}_1^e} =\;\not{\!\hat{P}_1}
=\;\not{\!p_3}\,+\!\not{\!\hat{p}_5}$.
Since the three-point amplitude is gauge-invariant,
we may pick the reference spinor $\ket{q}$ to our convenience.
In any case,
\be
   \frac{\sbraaket{5}{3}{q}}{\braket{\hat{5}\;\!q}}
    = \frac{\sbrasket{5}{3|4}{5}}{s_{45}-m^2} ,
\ee
which is evidently true for $\ket{q}=|4|5]$.
The residue~\eqref{grA12354compute}
is evaluated at the pole value of $z$ defined by
$\hat{s}_{35}-m^2=\sbraaket{5}{3}{\hat{5}}=0$.
This on-shell condition implies
\be
   z = \frac{m\sbraaket{5}{3}{5}}{\sbrasket{5}{3|4}{5}}, \qquad \quad
   \sket{\hat{4}^d} = \sket{4^d} - |5]
      \frac{\sbraaket{5}{3}{5} [5 4^d]}{\sbrasket{5}{3|4}{5}} , \qquad \quad
   \aket{\hat{5}} = -\frac{|3|5]\abrasket{5}{4}{5}}{[5|3|4|5]} ,
\ee
where in the last equation we have used the anticommutation relation
$|5|4|3|5] + |4|5|3|5] = 2(p_4\!\cdot p_5) |3|5]$.
Therefore, the numerator of the amplitude~\eqref{grA12354compute} becomes
\begin{align}
 & [5|3|4|5] \bigg(\:\!\!
         \braket{1 3} \sbraket{2 \hat{4}}
       + \braket{1 4} \sbraket{2 3}
       + \frac{1}{m} \braket{1 4} \sbraket{2 5} \braket{\hat{5}3}
       + (1 \leftrightarrow 2)\:\!\!\bigg) \\ &
    = \braket{1 3} \sbraket{2 4} [5|3|4|5]
       - \braket{1 3} \sbraket{2 5} \sbraaket{5}{3}{5} [5 4]
       + \braket{1 4} \sbraket{2 3} [5|3|4|5]
       - \braket{1 4} \sbraket{2 5}
         \sbraket{3 5} \abrasket{5}{4}{5}
       + (1 \leftrightarrow 2) . \nn
\end{align}
Here and below we omit the little-group indices of the external quarks
for the sake of brevity.
We may additionally use a Schouten identity
$-[24] [5|4|3|5] = [52] [4|4|3|5] + [45] [2|4|3|5]$,
which leads to our final expression for the five-point amplitude:
\beal
\label{eqA15}\!\!\!\!
  A(1^a\!,2^b\!,3^c\!,5^+\!,4^d)
   = \frac{-i}{s_{12} (s_{35}\!-\!m^2) (s_{45}\!-\!m^2)}
     \Big( \braket{14}\sbraket{23} \sbrasket{5}{3|4}{5}
       - \braket{14} \sbraket{25} \sbraket{35} (s_{45}\!-\!m^2) & \\
       +\,m \braket{13} \sbraket{25} \abrasket{4}{3}{5}
       + \braket{13} \sbrasket{2}{P_{45}|3}{5} \sbraket{45}
       + (1 \leftrightarrow 2) & \Big) .\!\!\!\!\!
\eeal
This formula is explicitly local because
we have obtained it from a single pole contribution.

%%%%%%%%%%%%%%%%%%%%%%%%%%%%%%%%%%%%%%%%%%%%%%%%%%%%
\paragraph{$\boldsymbol{n}$-point amplitude.}
We have chosen the above form of the five-point amplitude
because it can be seen as a special case
of an all-multiplicity formula that we have found.
To write it in a compact form,
let us introduce a shorthand notation for the propagator denominators,
such as $D_{3i \dots j} = P_{3i \dots j}^2 - m^2$,
as well as the following auxiliary massless spinors:
\begin{subequations} \begin{align}
   [c_i^j| & = [j| \prod_{k=j}^{i-1}
      \big\{\!\!\not{\!\!P}_{35 \dots k}\!\not{\!p}_{k+1}
           + D_{35 \dots k} \big\} , \qquad\:\:\quad j \leq i ,
\label{cspinor} \\
   |d_i^j] & =\!\prod_{k=i+1}^j\!
      \big\{\!\!\not{\!p}_{k-1}\!\not{\!\!P}_{4k \dots n}
           + D_{4k \dots n} \big\} |j] , \qquad \quad i \leq j ,
\label{dspinor}
\end{align} \end{subequations}
where the matrix factors are ordered by $k$ increasing from left to right.
These objects are slight generalizations of the spinors
appearing in the numerator of the two-quark formula~\eqref{QQggnAP},
although now they involve quark momenta $p_3$ and $p_4$.
Our all-multiplicity can then be written as
\begin{subequations}\label{eqAn1}
\begin{align}\!\!\!
   A&(1^a,2^b,3^c,5^+,\dots,n^+\!,4^d)
    = \frac{-i}{ s_{12} \prod_{j=5}^{n-1} \braket{j|j\!+\!1} } \Bigg\{
      \frac{1}{ \prod_{j=5}^n D_{4j \dots n} \abrasket{5}{3}{d_5^n} }
\nn \\* & \times\!
      \bigg[ \braket{14} \sbraket{23}
             \sbrasket{d_5^n}{3|P_{45 \dots n}}{d_5^n}
           - \braket{14} \sbraket{2|d_5^n} \sbraket{3|d_5^n}
             D_{45 \dots n}
           + m \braket{13} \sbraket{2|d_5^n} \abrasket{4}{3}{d_5^n}
\label{eqAn11} \\* & \qquad\:\!\!\quad
           + \braket{13} \sbrasket{2}{P_{45 \dots n}|3}{d_5^n}
             \bigg( \sbraket{4n} \prod_{j=5}^{n-1} D_{4j\dots n}
                  + m \sum_{i=5}^{n-1} \abrasket{4}{i}{d_{i+1}^n}
                      \prod_{j=5}^{i-1} D_{4j \dots n} \bigg)
      \bigg] \nn \\\!\!\!
       +&\;\!
      \sum_{i=6}^{n}
      \frac{ m \braket{i\!-\!1|i} [c_{i-1}^5 |d_i^n] }
           { \prod_{j=5}^{i-1} D_{35 \dots j}
             \prod_{j=i+1}^n D_{4j \dots n}
             \big(
             D_{35 \dots (i-1)} \abrasket{i\!-\!1}{P_{4i\dots n}}{d_i^n}
           + D_{4i\dots n} \abrasket{i\!-\!1}{P_{35 \dots (i-1)}}{d_i^n}
             \big) } \nn \\ &~\,\,\times\!
      \bigg[ \frac{ \abrasket{1}{P_{4i \dots n}}{d_i^n}
                    \sbraket{2|d_i^n} \braket{34} }
                  { \abrasket{i}{P_{4i \dots n}}{d_{i+1}^n} }
           + \frac{ \sbrasket{d_i^n}{P_{12}|P_{4i \dots n}}
                    {d_i^n} }
                  { D_{35 \dots i}
                    \abrasket{i}{P_{4(i+1) \dots n}}{d_{i+1}^n}
                  + D_{4(i+1) \dots n}
                    \abrasket{i}{P_{35 \dots i}}{d_{i+1}^n} }
\label{eqAn13} \\ & \qquad \qquad \qquad \qquad \qquad\;\:\quad
             \!\times\!
             \bigg( \braket{14} \sbraaket{2}{P_{12}}{3}
                  + \abrasket{1}{P_{4i \dots n}}{2} \braket{34}
                  + \frac{ \braket{1|i} \sbraket{2|d_i^n} \braket{34} }
                         { \abrasket{i}{P_{4i \dots n}}{d_{i+1}^n} }
             \bigg)
      \bigg]
      \Bigg\} \nn \\\!\!\!
       +&\: (1 \leftrightarrow 2 ) . \nn
\end{align} \end{subequations}
The last term in the sum~\eqref{eqAn13}
contains the auxiliary spinor $\sket{d_{n+1}^n}$,
which is not defined in by \eqn{dspinor},
because its lower index is larger than the upper index.
We supplement our definition by the following additional requirement:
\be
   \abrasket{n}{4}{d_{n+1}^{n}} = \abrasket{n}{P_{4n}}{d_{n+1}^{n}} = 1 .
\label{dspinorOver}
\ee
The only other spinor contraction appearing in the $i=n$ contribution,
$\abrasket{n}{P_{35 \dots n}}{d_{n+1}^n}$, is irrelevant,
as it comes with a vanishing prefactor $D_4=p_4^2-m^2$.
In this way, we were able to integrate this contribution into the general sum
instead of spelling it out separately.

Note that both $|c_i^j]$ and $|d_i^j]$ transform
under the massless little group of the spinor~$|j]$.
For instance, at five points the only such spinor is $|d_5^5]=|5]$,
because the sum~\eqref{eqAn13} does not yet contribute.
It is easy to see that the remaining term~\eqref{eqAn11}
produces precisely \eqn{eqA15}.
Having thus ensured the base case, we may proceed
to prove the complete $n$-point formula~\eqref{eqAn1} by induction.

%%%%%%%%%%%%%%%%%%%%%%%%%%%%%%%%%%%%%%%%%%%%%%%%%%%%
\paragraph{Inductive proof.}
We choose to use the BCFW shift $[4^d,n^+\rangle$,
for which each recursion step has exactly two contributions:
%\vspace{-5pt}
\begin{align}\label{eqAn1B}
   \includegraphics[valign=c,scale=0.9]{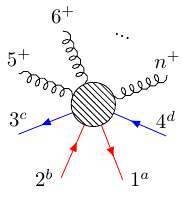} &
   \,=\,\includegraphics[valign=c,scale=0.9]{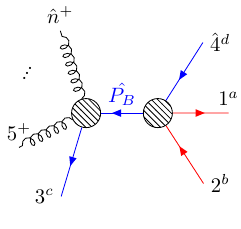}\,
    +\!\!\!\includegraphics[valign=c,scale=0.9]{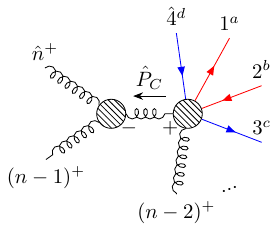}\,. \\*
   (A_n)~\,\qquad & \qquad\qquad\quad~\,(B_n)
   \qquad\qquad\qquad\qquad\quad~\,(C_n) \nn
\end{align}
The first diagram $B_n$ factorizes into
the four-quark seed amplitude~\eqref{eq4Fermion}
and the previously obtained \cite{Ochirov:2018uyq}
$n$-point amplitude~\eqref{QQggnAP} with two external quarks.
Thus we immediately get
\be
   B_n
    = \frac{ im \braket{3|{- \hat{P}_B^e}} [5| \prod_{j=5}^{n-2}
             ( \not{\!\!P}_{35 \dots j}\!\not{\!p}_{j+1}
                  + D_{35 \dots j} ) |n] }
           { s_{12} D_{35 \dots n} \prod_{j=5}^{n-1} D_{35 \dots j}
             \prod_{j=5}^{n-2} \braket{j|j\!+\!1} \braket{n-1|\hat{n}} }
      \Big( \braket{1|\hat{P}_{B,e}} \sbraket{2 \hat{4}}
          + \braket{1 4} \sbraket{2|\hat{P}_{B,e}}
          + (1 \leftrightarrow 2) \Big) ,
\ee
where we have taken care to separate the shifted factor
$\braket{n\!-\!1|\hat{n}}$ from the rest of the denominator.
We can also recognize the auxiliary spinor $[c_{n-1}^5|$
appearing in the unshifted numerator factor.
Similarly to the five-point calculation,
we need to insert the relevant pole values for the complex kinematics
(determined by $\hat{D}_{35 \dots n} = 0$), for which we obtain
\be
   z_B = \frac{m D_{35 \dots n}}{[n|P_{35 \dots n}|4|n]} , \qquad \quad
   \ket{\hat{n}} = \frac{ |P_{4n}|n]D_{35 \dots(n-1)}
                        + |P_{35 \dots(n-1)}|n] D_{4n} }
                        { [n|4|P_{35 \dots n}|n] } . %\qquad \quad
%   |\hat{4}^d] = |\hat{4}^d]
%    + \frac{|n] [n 4^d] D_{35 \dots n}}{[n|4|P_{35 \dots n}|n]} ,
\label{eq1polB}
\ee
We also need to eliminate the internal momentum $\hat{P}_B$:
\beal \label{eq:internalPB}
   \braket{3|{-\hat{P}_B^e}}
   \Big( \braket{1|\hat{P}_{B,e}} [2 \hat{4}]
       + \braket{1 4} [2|\hat{P}_{B,e}] \Big)
    = m \braket{1 3} [2 \hat{4}]
    - \braket{1 4} [2|P_{12\hat{4}}\ket{3} & \\
    =-\braket{1 4} [2|P_{12}\ket{3} - \bra{1}4|2] \braket{34}
    - \frac{D_{35 \dots n}}{[n|4|P_{35 \dots n}|n]}
      \bra{1}4|n] [2 n] \braket{3 4} & ,
\eeal
where we have inserted the correspondingly shifted expression
for $|\hat{4}]$ and further simplified the result.
Therefore, the first residue in \eqn{eqAn1B} becomes
\begin{align}\!\!\!
   B_n =\,& \frac{-i m}{ s_{12} \prod_{j=5}^n D_{35 \dots j}
                         \prod_{j=5}^{n-2} \braket{j|j\!+\!1} }
      \frac{ [c_{n-1}^5|n] }
           { D_{35 \dots (n-1)} \abrasket{n\!-\!1}{P_{4n}}{n}
           + D_{4n} \abrasket{n\!-\!1}{P_{35 \dots (n-1)}}{n} }\!\\* & \times\!
      \bigg[ [n|4|P_{35 \dots n}|n]
             \Big( \braket{1 4} [2|P_{12}\ket{3}
                 + \bra{1}4|2] \braket{34} \Big)
           + D_{35 \dots n} \bra{1}4|n] [2 n] \braket{3 4} \bigg]
    + (1 \leftrightarrow 2) . \nn
\end{align}
After slight rearrangement, this can be seen to match
the $i=n$ contribution in the sum~\eqref{eqAn13} of the $n$-point formula.

The second residue $C_n$ is where we really need an inductive argument,
as it factorizes into an $(n-1)$-point amplitude of the same type
that we aim to compute:
\be
   C_n = A((n\!-\!1)^+\!,\hat{n}^+\!,-\hat{P}_C^-) \frac{-i}{s_{(n-1)n}}
   A(1^a,2^b,3^c,5^+\!,\dots,(n\!-\!2)^+\!,\hat{P}_C^+,\hat{4}^d) .
\label{eqAn1C}
\ee
By the induction hypothesis, we may express
the right-hand side of the factorization using the formula~\eqref{eqAn1},
in which the complex kinematics is determined by the pole
$\hat{s}_{(n-1)n}=0$:
\be\!\!
   z_C = \frac{m \braket{n\!-\!1|n}}{\bra{n\!-\!1}4|n]} , \qquad~
   \ket{\hat{n}} = \ket{n\!-\!1}
      \frac{D_{4n}}{\bra{n\!-\!1}4|n]} , \qquad~
   [\hat{4}^d| = [4^d|
    + \frac{[4^d n] \braket{n\!-\!1|n}}{\bra{n\!-\!1}4|n]} [n| .
\label{eq1polC}
\ee
The internal momentum $\hat{P}_C$ may then be decomposed into
\be
   \ket{\hat{P}_C} = \ket{n\!-\!1} , \qquad \quad
   |\hat{P}_C] = |n\!-\!1] + \frac{D_{4n} |n]}{\bra{n\!-\!1}4|n]}
%    = \frac{|n\!-\!1|P_{4n}|n] + D_{4n} |n]}{\bra{n\!-\!1}4|n]}
    = \frac{|d_{n-1}^n]}{\bra{n\!-\!1}4|n]} ,
\label{InternalMomentumC}
\ee
up to a rescaling,
where we have recognized the first appearance
of the auxiliary spinor $|d_{n-1}^n]$, defined in \eqn{dspinor}.
The non-recursive part of the residue can be identified with the factor
\be
   \frac{-i}{s_{(n-1)n}} A((n\!-\!1)^+\!,\hat{n}^+\!,-\hat{P}_C^-)
%    = \frac{-[n\!-\!1|n]^3}{s_{(n-1)n} [\hat{P}_C|n\!-\!1] [n|\hat{P}_C]}
%    = \frac{ [n\!-\!1|n]^2 \bra{n\!-\!1}P_{4n}|n]^2 }
%           { \braket{n\!-\!1|n} D_{4n} [n|n\!-\!1] [n|n\!-\!1|4|n] }
    = \frac{\bra{n\!-\!1}4|n]}{D_{4n} \braket{n\!-\!1|n}} .
\label{eqsubA3}
\ee
Combining it together with the prefactor in the $(n-1)$-point amplitude,
we can already observe the formation of the $n$-point prefactor
\be
   \frac{\bra{n\!-\!1}4|n]}{D_{4n} \braket{n\!-\!1|n}} \times
   \frac{-i}{ s_{12} \prod_{j=5}^{n-3} \braket{j|j\!+\!1}
              \braket{n\!-\!2|\hat{P}_C} }
 = \frac{-i}{s_{12} \prod_{j=5}^{n-1} \braket{j|j\!+\!1}}
   \frac{\bra{n\!-\!1}4|n]}{D_{4n}} .
\label{eqAn1prefactor}
\ee

Let us now inspect how the recursion affects
the terms of the formula~\eqref{eqAn1} in the curly brackets.
We immediately notice that almost all of the featured momentum sums
involve both $\hat{p}_4$ and $\hat{P}_C$
and so can be easily reduced to an unshifted form.
Namely,
\be
   P_{4j \dots n} ~\to~
      \hat{P}_{4j \dots (n-2)P_C} = P_{4j \dots n}, \qquad  \qquad
   D_{4j \dots n} ~\to~
      \hat{D}_{4j \dots (n-2)P_C} = D_{4j \dots n} ,
\label{eqAn13step1}
\ee
where the arrows correspond to going from the $n$-point expressions
to the shifted $(n-1)$-point expressions on the right-hand side
of the residue~\eqref{eqAn1C}.
From a similar examination of the auxiliary spinors,
we see that $[c_{i-1}^5|$ stay entirely unshifted,
whereas $|d_i^n]$ become
\be
   |d_i^n] ~\to~ |\hat{d}_i^{P_C}]
    =\!\prod_{k=i+1}^{n-1}\!
      \big\{\!\!\not{\!p}_{k-1}\!\not{\!\!P}_{4k \dots n}
           + D_{4k \dots n} \big\} \frac{|d_{n-1}^n]}{\bra{n\!-\!1}4|n]}
    = \frac{|d_i^n]}{\bra{n\!-\!1}4|n]} .
\label{eqAn13step2}
\ee
Applying these replacements to the first contribution~\eqref{eqAn11}
in the curly brackets, we find
\small
\begin{align} &\!\!\!
   \frac{ \bra{n\!-\!1}4|n] }
        { \prod_{j=5}^{n-1} D_{4j \dots n} \bra{5}3|d_5^n] }
      \bigg[ \braket{14} [23]
             \frac{[d_5^n|3|P_{45 \dots n}|d_5^n]}{\bra{n\!-\!1}4|n]^2}
           - \braket{14}
             \frac{[2|d_5^n] [3|d_5^n]}{\bra{n\!-\!1}4|n]^2}
             D_{45 \dots n}\!
           + m \braket{13}
             \frac{[2|d_5^n] \bra{4}3|d_5^n]}{\bra{n\!-\!1}4|n]^2} \nn
             \\* & \!\!\:\qquad \qquad \qquad \qquad
           + \braket{13}
             \frac{[2|P_{45 \dots n}|3|d_5^n]}{\bra{n\!-\!1}4|n]}
             \bigg( [\hat{4}\hat{P}_C]
                    \prod_{j=5}^{n-2} D_{4j \dots n}
                  + m \sum_{i=5}^{n-2}
                    \frac{\bra{4}i|d_{i+1}^n]}{\bra{n\!-\!1}4|n]}
                    \prod_{j=5}^{i-1} D_{4j \dots n} \bigg)
      \bigg]
\label{eqAn11derivation} \\ & \!\!\!
 = \frac{ ^{D_{4n}}\!/\!_{\bra{n-1}4|n]} }
        { \prod_{j=5}^n D_{4j \dots n} \abrasket{5}{3}{d_5^n} }
      \bigg[ \braket{14} \sbraket{23}
             \sbrasket{d_5^n}{3|P_{45 \dots n}}{d_5^n}
           - \braket{14} \sbraket{2|d_5^n} \sbraket{3|d_5^n} D_{45 \dots n}
           + m \braket{13} \sbraket{2|d_5^n} \abrasket{4}{3}{d_5^n}
             \nn \\* & \!\!\quad
           + \braket{13} \sbrasket{2}{P_{45 \dots n}|3}{d_5^n}
             \bigg(\!\Big( D_{4(n-1)n} [4n] + m \bra{4}n\!-\!1|n] \Big)
                    \prod_{j=5}^{n-2} D_{4j \dots n}
                  + m \sum_{i=5}^{n-2} \abrasket{4}{i}{d_{i+1}^n}
                    \prod_{j=5}^{i-1} D_{4j \dots n} \bigg)
      \bigg] , \nn
\end{align}
\normalsize
where we have used the identity
\be
   [\hat{4}^d \hat{P}_C] = \frac{1}{\bra{n\!-\!1}4|n]}
      \Big( D_{4(n-1)n} [4^d n] + m \bra{4^d}n\!-\!1|n] \Big) .
\ee
After combining \eqn{eqAn11derivation}
with the prefactor~\eqref{eqAn1prefactor}
and slightly rearranging the last line,
we retrieve the precise form of the first term~\eqref{eqAn11}
in the $n$-point expression.
In other words, this term is converted into itself
under the recursion~\eqref{eqAn1B}.

We now turn to the part the shifted $(n-1)$-point amplitude
which involves the sum~\eqref{eqAn13}.
First, we isolate the last term from the rest of the sum,
in which we use
$\bra{n}4|d_{n+1}^n] \to \bra{\hat{P}_C}4|\hat{d}_{P_C+1}^{P_C}] = 1$
in line with the supplementary definition~\eqref{dspinorOver},
which simplifies this contribution.
Together with the main sum, it gives
\begin{align}
 &\;\!\sum_{i=6}^{ n-2}
   \frac{ m \braket{i\!-\!1|i} [c_{i-1}^5 |d_i^n] }
        { \prod_{j=5}^{i-1} D_{35 \dots j}
            \prod_{j=i+1}^{n-1}\!D_{4j \dots n}
            \big(
            D_{35 \dots (i-1)} \abrasket{i\!-\!1}{P_{4i\dots n}}{d_i^n}
          + D_{4i\dots n} \abrasket{i\!-\!1}{P_{35 \dots (i-1)}}{d_i^n}
            \big)\!}\!
        \nn \\* &\!\times\!\frac{1}{\bra{n\!-\!1}4|n]}
        \bigg[ \frac{ \abrasket{1}{P_{4i \dots n}}{d_i^n}
                      \sbraket{2|d_i^n} \braket{34} }
                    { \abrasket{i}{P_{4i \dots n}}{d_{i+1}^n} }
             + \frac{ \sbrasket{d_i^n}{P_{12}|P_{4i\dots n}}{d_i^n} }
                    { D_{35 \dots i}
                      \abrasket{i}{P_{4(i+1) \dots n}}{d_{i+1}^n}
                    + D_{4(i+1) \dots n}
                      \abrasket{i}{P_{35 \dots i}}{d_{i+1}^n}} \nn
        \\* & \qquad \qquad \qquad \qquad \qquad \qquad \qquad~\;\times\!
        \bigg( \braket{14} \sbraaket{2}{P_{12}}{3}
        + \abrasket{1}{P_{4i \dots n}}{2} \braket{34}
        + \frac{ \braket{1|i} \sbraket{2|d_i^n} \braket{34} }
        { \abrasket{i}{P_{4i \dots n}}{d_{i+1}^n} }
        \bigg) \bigg] \nn \\
 & + \frac{ m \braket{n\!-\!2|n\!-\!1} [c_{n-2}^5 |d_{n-1}^n] }
        { \prod_{j=5}^{n-2} D_{35 \dots j}
          \big(
          D_{35 \dots (n-2)} \abrasket{n\!-\!2}{P_{4(n-1)n}}{d_{n-1}^n}
        + D_{4(n-1)n} \abrasket{n\!-\!2}{P_{35 \dots (n-2)}}{d_{n-1}^n}
          \big)\!}\! \nn
      \\* & \quad \times\!\frac{1}{\bra{n\!-\!1}4|n]^2}
      \bigg[ { \abrasket{1}{P_{4(n-1)n}}{d_{n-1}^n}
                    \sbraket{2|d_{n-1}^n} \braket{34} }
           + \frac{ \sbrasket{d_{n-1}^n}{P_{12}|P_{4(n-1)n}}{d_{n-1}^n} }
                  { \hat{D}_{35 \dots P_C} }
\label{eqAn13derivation}
      \\* & \qquad \qquad \qquad \quad\:\,\times\!
      \bigg( \braket{14} \sbraaket{2}{P_{12}}{3}
           + \abrasket{1}{P_{4(n-1)n}}{2} \braket{34}
           + \braket{1|n\!-\!1} \frac{\sbraket{2|d_{n-1}^n}}{\bra{n\!-\!1}4|n]}
             \braket{34}
      \bigg) \bigg] , \nn
\end{align}
where we have also taken into account
the transition rules~\eqref{eqAn13step1} and \eqref{eqAn13step2}.
The only quantity that still needs to be evaluated on the shift kinematics is
\begin{align}
    \hat{D}_{35 \dots P_C} =
    \frac{ D_{35 \dots (n\!-\!1)} \abrasket{n\!-\!1}{P_{4 n}}{d_{n}^n}
         + D_{4n} \abrasket{n\!-\!1}{P_{35 \dots (n\!-\!1)}}{d_{n}^n}}
         {\bra{n\!-\!1}4|n]} .
\end{align}
Here we have replaced $|n]=|d_n^n]$
so as to expose the similarity to the summands in the main sum above.
We can therefore integrate the last three lines of \eqn{eqAn13derivation}
into this sum as its $i=n-1$ contribution.
If we multiply this sum by the prefactor~\eqref{eqAn1prefactor}
and add the $i=n$ term arising from by the first residue~$B_n$,
the result will coincide with the corresponding sum
in the all-multiplicity amplitude~\eqref{eqAn1}.
This concludes our proof.

%%%%%%%%%%%%%%%%%%%%%%%%%%%%%%%%%%%%%%%%%%%%%%%%%%%%
\subsection{Gluon insertions between distinctly flavored quarks}
\label{sec:AmpType2}
%%%%%%%%%%%%%%%%%%%%%%%%%%%%%%%%%%%%%%%%%%%%%%%%%%%%

Let us now proceed to presenting results for the other family
of color orderings of the form $A(1^a,2^b,3^c,4^d,5^+,\dots ,n^+)$,
where all gluons are inserted between quarks $1$ and $4$
that carry different flavors. We will follow a similar path as the previous subsection.

%%%%%%%%%%%%%%%%%%%%%%%%%%%%%%%%%%%%%%%%%%%%%%%%%%%%
\paragraph{5-point amplitude.}
At five points, the $[4^d,5\rangle$ shift~\eqref{Shift45}
now implies two BCFW residues,
the second of which contains two helicity configurations:
%\vspace{-5pt}
\begin{align} \!\!\!\!\!
   \includegraphics[valign=c,scale=0.9]{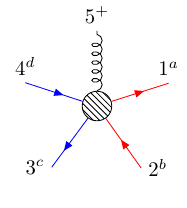}\! &=
      \includegraphics[valign=c,scale=0.9]{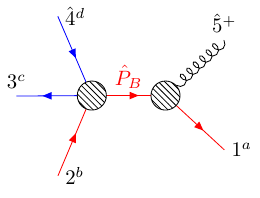}\!\!+\!
      \includegraphics[valign=c,scale=0.9]{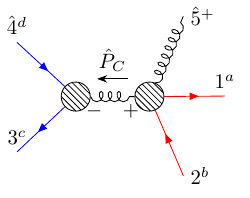}\!+\!\!
      \includegraphics[valign=c,scale=0.9]{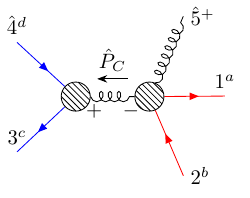}\!\!\!\!. \nn \\
   (A_5) \quad\quad~ & \qquad\qquad\quad~\, (B_5) \qquad\qquad\qquad\qquad~
   (C_5) \qquad\qquad\qquad\qquad~~ (E_5)
\label{eq:diaA52}
\end{align}
From this recursion, the amplitude can be derived in the following form:
\begin{align}
   A(1^a, 2^b,3^c,4^d,5^+)
    = \frac{-i}{[5|1|P_{12}|3|P_{12}|5]}
      \bigg\{ \frac{M \braket{12} [5|3|P_{12}|5]}{s_{12} s_{34}}
              \big( [35] \bra{4}P_{12}|5]
                  + [45] \bra{3}P_{12}|5] & \big) \nn \\
            + \frac{m \braket{34} [5|1|2|5]}{s_{12} s_{34}}
              \big( [15] \bra{2}P_{12}|5]
                  + [25] \bra{1}P_{12}|5] \big)
            - \frac{[5|1|4|5]}{D_{15} D_{45}}
              \Big[ D_{15} [45]
                    \big( \braket{13} [25] + \braket{23} [15] \big) & \nn \\
                  + D_{45} [15]
                    \big( \braket{24} [35] + \braket{23} [45] \big)
                  + [5|1|4|5]
                    \big( \braket{13} [24] + \braket{14} [23]
                        + [13] \braket{24} + [14] \braket{23} \big) &
              \Big]
      \bigg\}
\label{eqA52}
\end{align}
Note that the mass in the denominators is understood to depend
on the relevant flavor,
\eg $D_{15} = P_{15}^2 - M^2$ and $D_{45} = P_{45}^2 - m^2$.
We follow this convention throughout this article.
A detailed derivation can be found in \app{sec:proofeqA52}.

%%%%%%%%%%%%%%%%%%%%%%%%%%%%%%%%%%%%%%%%%%%%%%%%%%%%
\paragraph{$\boldsymbol{n}$-point amplitude.}
Let us now consider the all-multiplicity amplitude
with all gluons appearing between the distinctly flavored quarks $1$ and $4$,
namely $A(1^a,2^b,3^c,4^d,5^+,\dots,n^+)$.
The massive-massless shift $[4^d,5\rangle$ now results in four diagrams:
%\vspace{-5pt}
\begin{equation}\label{eqnptdia2}
\begin{aligned}
   \includegraphics[valign=c]{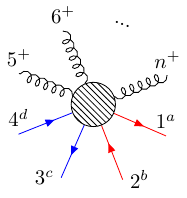} = &\,
   \includegraphics[scale=0.95,valign=c]{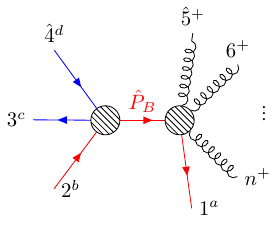} +
   \includegraphics[scale=0.95,valign=c]{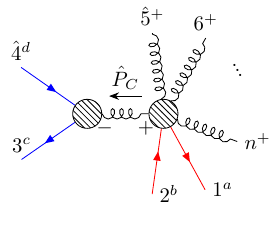} \\
   (A_n) \qquad\qquad\, & \quad\qquad\qquad\,
   (B_n) \quad\qquad\qquad\qquad\qquad\quad (C_n) \\ & +\!
   \includegraphics[scale=0.95,valign=c]{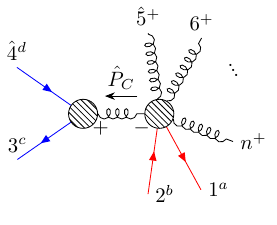}\!+
   \includegraphics[scale=0.95,valign=c]{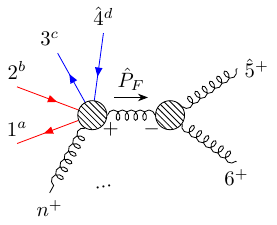} . \\ &
   \qquad\qquad\quad\, (E_n) \qquad\qquad\qquad\qquad\qquad\quad\! (F_n)
\end{aligned}
\end{equation}
Note that the contributions $B_n$, $C_n$, and $E_n$
have already appeared at five points,
whereas $F_n$ is a new residue appearing for $n>5$,
and it is where the main recursion happens.
Let us proceed directly to the final expression,
which we again write in terms of auxiliary spinors
that originate from the numerator of the two-quark formula~\eqref{QQggnAP}:
\begin{subequations} \begin{align}
   |a_i^j] & =\!\prod_{k=i+1}^{j}\!
      \big\{\!\!\not{\!p}_{k-1}\!\not{\!\!P}_{1k \dots n}
           + D_{1k \dots n} \big\} |j] , \qquad \quad i \leq j ,
\label{aspinor} \\
   [d_i^j| & = [j| \prod_{k=j}^{i-1}
      \big\{\!\!\not{\!\!P}_{45 \dots k}\!\not{\!p}_{k+1}
           + D_{45 \dots k} \big\} , \qquad\:\:\quad j \leq i ,
\label{dspinor2} \\
   |e_i^d] & = \prod_{k=6}^i D_{45 \dots k} |4^d]
    - m \sum_{j=6}^i \prod_{k=5}^{j-2}
      \big\{\!\!\not{\!\!P}_{45 \dots k}\!\not{\!p}_{k+1}
           + D_{45 \dots k} \big\}
      \bigg[ \prod_{k=j+1}^i D_{45 \dots k} \bigg]
      \!\not{\!p}_j \ket{4^d} .
\label{espinor}
\end{align} \end{subequations}
Note that $|a_i^j]$ carries the helicity information of gluon $j$,
$|d_i^j]$ carries that of $j$,
and $|e_i]$ inherits the little-group index $d$ from quark $4$.
For the $i=5$ terms in the sums,
we formally define $|d_4^5]$ by imposing
\begin{equation} \label{eqspinord2}
    \bra{5}4|d_4^5] = 1.
\end{equation}
It also assures that the five-point amplitude~\eqref{eqA52}
is covered by the following $n$-point formula:
\newpage
\small
\begin{subequations} \label{eqAn2}
\begin{align}
   A&(1^a, 2^b,3^c,4^d,5^+,\dots ,n^+)
 = \frac{\!-i}{\prod_{j=5}^{n-1} \braket{j|j\!+\!1}}
   \Bigg\{
   \frac{ 1 }{ \prod_{j=5}^{n-1} D_{4\dots j}
               [d_n^5|P_{12}|3|P_{12}|1|d_n^5] } \nn \\* &\:\!\!\times\!
   \Bigg[ \frac{ [e_n|5] [d_n^5|1|P_{123}|d_n^5] }
               { D_{123} \bra{n}P_{123}|d_{n-1}^5] }
          \Big( \braket{13} [2|d_n^5] + \braket{23} [1|d_n^5] \Big)
        + \frac{1}{s_{12} \bra{n}P_{12}|3|P_{12n}|d_{n-1}^5]} \nn \\* &
          \quad\;\times\!
          \bigg[ M \braket{12} [d_n^5|P_{12}|3|d_n^5]
                 \Big( [3|d_n^5] \bra{4}P_{12}|d_n^5]
                     + [e_n|5] \bra{3}P_{12}|d_n^5]
                 \Big) \nn \\* & \qquad\;\;\;
               + m \braket{34} [d_n^5|2|1|d_n^5]
                 \Big( [1|d_n^5] \bra{2}P_{12}|d_n^5]
                     + [2|d_n^5] \bra{1}P_{12}|d_n^5]
                 \Big) \bigg]
\label{eqAn21} \\ &~\quad
        + \frac{ [d_n^5|1|P_{123}|d_n^5] }
               { D_{1n} \bra{n}P_{123n}|d_{n-1}^5]
                - D_{123n} \bra{n}P_{1n}|d_{n-1}^5] }
          \bigg[ \braket{24} [1|d_n^5] [3|d_n^5]
               + \braket{23} [1|d_n^5] [e_n|5] \nn \\ & \qquad\;\;\;
               - \frac{[d_n^5|1|P_{123}|d_n^5]}{D_{123}}
                 \Big( [13] \braket{24} + \braket{14} [23] \Big)
               + \frac{ [d_n^5|1|P_{123}|d_n^5] }
                      { D_{123}^2 \bra{n}P_{123}|d_{n-1}^5] } \nn \\* &
                 \qquad~~\,\quad \times\!
                 \Big( \braket{13} [2|P_{123}\ket{n} [e_n|5]
                     + \braket{23} [1|P_{123}\ket{n} [e_n|5]
                     - m \braket{13} \braket{4n} [2|d_n^5]
                     - m \braket{23} \braket{4n} [1|d_n^5]
                 \Big)
          \bigg]
   \Bigg]
\nn \\
+& \sum_{i=5}^{n-1}
   \frac{ M \braket{i|i\!+\!1}  [a_{i+1}^n|d_i^5]
          [d_i^5|P_{23}|P_{4\dots i}|d_i^5] }
        { [d_i^5|P_{3\dots i}|2|3|P_{3\dots i}|d_i^5]
          \prod_{l=i}^{n-1} D_{2\dots l} \prod_{k=5}^{i-1} D_{4\dots k}
           \big( D_{4\dots i} \bra{i+1}P_{2\dots i}|d_i^5]
               - D_{2\dots i} \bra{i+1}P_{4\dots i}|d_i^5]
           \big) } \nn \\* &\:\!\!\times\!
   \Bigg[ \frac{ [d_i^5|P_{23}|P_{4\dots i}|d_i^5] }
               { D_{4\dots (i-1)} \bra{i}P_{2\dots (i-1)}|d_{i-1}^5]
               - D_{2\dots (i-1)} \bra{i}P_{4\dots (i-1)}|d_{i-1}^5] }
          \bigg[ \bra{1}P_{2\dots i}|3] \braket{24}
               + M \braket{14} [23] \nn \\* & \qquad\;\;\;
                - \frac{1}{D_{4 \dots i} \bra{i}P_{4\dots (i-1)}|d_{i-1}^5]}
                  \Big( \bra{1}P_{23}|P_{4\dots i}\ket{i} \braket{23} [e_i|5]
                       - D_{4\dots i} \braket{1i} \braket{24} [3|d_i^5]
\label{eqAn22} \\* & \qquad~~\,\quad
                       + m \bra{1}P_{2\dots i}|d_i^5] \braket{23} \braket{4i}
                       + M \braket{13} \bra{i}P_{4\dots i}|2] [e_i|5]
                       + m M \braket{13} [2|d_i^5] \braket{4i}
                       + m^2 \braket{1i} \braket{23} [e_i|5]
                  \Big)
           \bigg] \nn \\* &~\quad
         + \frac{1}{\bra{i}P_{4\dots (i-1)}|d_{i-1}^5]}
           \bigg[ \bra{1}P_{23}|d_i^5] \braket{23} [e_i|5]
                - \bra{1}P_{4\dots i}|d_i^5] [3|d_i^5] \braket{24}
                + M \braket{13} [2|d_i^5] [e_i|5]
           \bigg]
    \Bigg]
\nn \\
+& \sum_{i=5}^{n-1}
   \frac{ M \braket{12} \braket{i|i\!+\!1}
         \big( [3|d_i^5] \bra{4}P_{3\dots i}|d_i^5]
             + \bra{3}P_{3\dots i}|d_i^5] [e_i|5]
         \big) [d_i^5|P_{3\dots i}|3|d_i^5] }
        { s_{3\dots i} \prod_{l=i}^{n-1}\!D_{2\dots l}
          \prod_{k=5}^{i-1} D_{4\dots k}
          \bra{i\!+\!1}P_{3\dots i}|3|P_{3\dots i}|d_i^5]
          \bra{i}P_{3\dots (i-1)}|3|P_{3\dots (i-1)}|d_{i-1}^5] }
\label{eqAn23} \\* & \quad\!\times\!
   \frac{ [a_{i+1}^n|P_{3\dots i}|2|3|P_{3\dots i}|d_i^5] }
        { [d_i^5|P_{3\dots i}|2|3|P_{3\dots i}|d_i^5] }
\nn \\
+& \sum_{i=5}^{n-1}
   \frac{ m \braket{34} \braket{i|i\!+\!1} }
        { s_{3\dots i} \prod_{k=5}^{i-1} D_{4\dots k}
           \bra{i}P_{3\dots (i-1)}|3|P_{3\dots (i-1)}|d_{i-1}^5]
         } \nn \\* & \:\!\!\times\!
   \Bigg[ \frac{ [d_i^5|P_{3\dots i}|2|1|P_{3\dots i}|d_i^5]
                 \big( \bra{1}P_{3\dots i}|d_i^5] [2|P_{12}|P_{3\dots i}|d_i^5]
                     + \bra{2}P_{3\dots i}|d_i^5] [1|P_{12}|P_{3\dots i}|d_i^5]
                 \big) }
               { s_{12} \bra{n}P_{12}|2|P_{3\dots i}|d_i^5]
                 \bra{i\!+\!1}P_{3\dots i}|3|P_{3\dots i}|d_i^5] } \nn \\* & \quad\;
        + \frac{ M s_{3\dots i} [d_i^5|2|P_{3\dots i}|d_i^5] [a_{i+1}^n|d_i^5]
                \big( \braket{12} [d_i^5|P_{3\dots i}|2|d_i^5]
                    + \bra{1}P_{3\dots i}|d_i^5] \bra{2}P_{3\dots i}|d_i^5]
                \big) }
               { \prod_{j=i}^{n-1}\!D_{2\dots j}
                 \bra{i\!+\!1}P_{3\dots i}|2|P_{3\dots i}|d_i^5]
                 [d_i^5|P_{3\dots i}|2|3|P_{3\dots i}|d_i^5] }
\label{eqAn24} \\* & \quad\;
       +\!\sum_{k=i+1}^{n-1}
          \frac{ M \braket{k|k\!+\!1}
                 [d_i^5|P_{3\dots i}|2|P_{3\dots k}|P_{3\dots i}|d_i^5]
                 [a_{k+1}^n|P_{3\dots k}|P_{3\dots i}|d_i^5] }
               { s_{3\dots k} \prod_{j=k}^{n-1} D_{2\dots j}
                 \bra{k}P_{3\dots k}|2|P_{3\dots i}|d_i^5]
                 \bra{k\!+\!1}P_{3\dots k}|2|P_{3\dots i}|d_i^5]
                 \bra{i\!+\!1}P_{3\dots i}|3|P_{3\dots i}|d_i^5] } \nn \\* &
\qquad~\:\,\quad \times\!
          \Big( \braket{12} [d_i^5|P_{3\dots i}|2|P_{3\dots k}|P_{3\dots i}|d_i^5]
               + s_{3\dots k} \bra{1}P_{3\dots i}|d_i^5]
                 \bra{2}P_{3\dots i}|d_i^5] \Big)
   \Bigg]
   \Bigg\} . \nn
\end{align} \end{subequations}
\normalsize
Here the momentum sums like $P_{3\dots i}$ are always understood
in the ``clockwise'' fashion, namely
$P_{3\dots i} = p_3 + p_4 + \ldots + p_{i-1} + p_i$.
The proof of \eqn{eqAn2} is very similar to that in the previous subsection,
and we leave it to \app{sec:proofeqAn2}.

%%%%%%%%%%%%%%%%%%%%%%%%%%%%%%%%%%%%%%%%%%%%%%%%%%%%
\subsection{Gluons in between both distinctly and  like-flavored quarks}
\label{sec:AmpType3}
%%%%%%%%%%%%%%%%%%%%%%%%%%%%%%%%%%%%%%%%%%%%%%%%%%%%

Recall that in order to construct
the full color-dressed amplitude with four quarks
one needs \cite{Johansson:2015oia}
color-ordered amplitudes of the general form
$A({\color{red} \u{1}},{\color{red} \o{2}},{\color{blue} \u{3}},
   \sigma({\color{blue} \o{4}},5,\dots,n))$,
This of course includes the cases
where gluons are inserted on both sides of quark $4$.
Due to the complexity of the recursive structure,
we refrain from computing the closed formula in this ordering.
Alternatively, here we present a BCFW shift
to extract the numerical result of the amplitude.
In this ordering, we found it most efficient to use the shift $[3^c, 5^+\rangle$. The factorization channels are
%\vspace{-5pt}
\begin{equation}
\begin{split}\label{eq:3diag}
      \includegraphics[scale=0.93, valign=c]{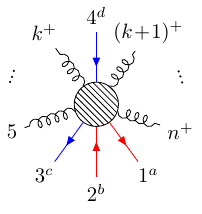} = &
      \includegraphics[scale=0.93, valign=c]{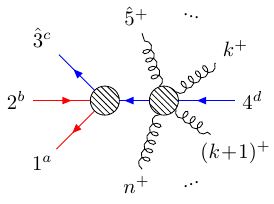}
      + \sum_{l=k+1}^n
      \includegraphics[scale=0.93, valign=c]{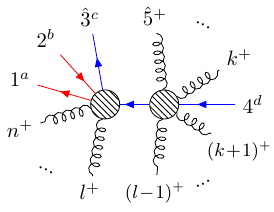}  \\
      &+ \includegraphics[scale=0.95, valign=c]{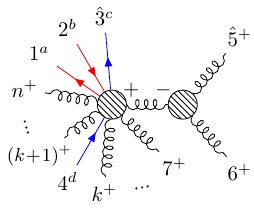}
\end{split}
\end{equation}
On the right-hand side of the first line,
we have the previously obtained four-quark amplitudes of form~\eqref{eqAn2},
as well as two-quark amplitudes with gluons in arbitrary color orderings
--- as opposed to the closed formula~\eqref{QQggnAP},
in which the two quarks are color-adjacent.
General orderings may, however, be still computed from \eqn{QQggnAP}
using the Kleiss-Kuijf relation
\be\!\!
    A(3^c, 5^+\!, \dots, k^+\!, 4^d, (k\!+\!1)^+\!, \dots, l^+) =
    (-1)^{l-k} \sum_{\sigma}
    A(3^c, \sigma(5)^+\!, \sigma(6)^+\!, \dots, \sigma(l)^+\!, 4^d) ,
\ee
where the sum is over the shuffle of the gluons $\sigma \in \{5,6,\dots, k\} \shuffle \{l, (l\!-\!1), \dots, (k\!+\!1)\}$.

The last term in \eqn{eq:3diag} involves
a lower-point version of the amplitude on the right-hand side,
with gluons 5 and 6 replaced by a single gluon with a complex momentum.
This is where the recursion step takes place, unless $k=5$,
in which case this term vanishes, and the amplitude
is obtained from the previously computed amplitudes.

%%%%%%%%%%%%%%%%%%%%%%%%%%%%%%%%%%%%%%%%%%%%%%%%%%%%
\section{Numerical evaluation}
\label{sec:numerical}
%%%%%%%%%%%%%%%%%%%%%%%%%%%%%%%%%%%%%%%%%%%%%%%%%%%%

In this section we analyze our analytic results from the point of view of
the time needed to evaluate them numerically on a computer.

The main benchmark for us will be our own implementation
of the off-shell Berends-Giele (BG) recursion~\cite{Berends:1987me},
also in C++.
This method essentially provides an inductive realization
of conventional Feynman diagrams and is therefore very flexible
and relatively easy to implement.
In particular, its validity does not depend on such properties
of the theory under consideration
as the boundary behavior for large complex-shifted momenta.
Furthermore, in the case of massless gauge theory
it was found \cite{Dinsdale:2006sq,Badger:2012uz}
to be more efficient than the on-shell BCFW recursion,
in the case of generic helicities of the external particles.
For these reasons, the BG recursion
is the method of choice for numerical evaluation
of tree-level amplitudes for phenomenological purposes
\cite{Duhr:2006iq,Giele:2008bc,Ellis:2008qc,Lazopoulos:2008ex,Badger:2010nx}.

It is perhaps somewhat counter-intuitive
that closed analytic formulae even for massless QCD amplitudes
are not always able to outperform the BG recursion.
Indeed, \rcite{Badger:2012uz} observed that the evaluation time
of analytic formulae for massless QCD amplitudes~\cite{Dixon:2010ik}
(which were obtained
by solving~\cite{Drummond:2008cr} the on-shell BCFW recursion
in ${\cal N}=4$ supersymmetric Yang-Mills theory)
grows faster than the time needed to run
a numerically efficient implementation of the BG recursion
already at the N$^2$MHV level of analytic complexity.
Recall that massless amplitudes possess a ``complexity-peeling'' property,
which is reflected by the fact that the color-ordered N$^k$MHV amplitude
formulae involve $k$ nested sums.

This ``complexity-peeling'' property is scrambled in the massive case.
Indeed, in the massive spinor-helicity formalism a single amplitude
\eg with $4$ massive quarks constitutes a $2 \times 2 \times 2 \times 2$ matrix,
each element of which can be associated with a massless-quark
helicity amplitude in the high-energy limit.
So it is natural that the complexity of the analytic formula
in the massive case can be no simpler than that of
the most complicated massless helicity configuration that it incorporates.
In fact, we observe that the massive formulae
are generally even more complicated than this naive expectation,
as most easily seen in the two-quark case.
For instance, the ``one-minus'' amplitude \cite{Ochirov:2018uyq}
\beal
   A(\u{1}^a,3^-\!,4^+\!,\dots,n^+\!,\o{2}^b)
    = \frac{\!-i}{\prod_{j=3}^{n-1} \braket{j|j\!+\!1}} \bigg[
      \frac{\braket{3|1|2|3}}{s_{12} \braket{3|1|P_{12}|n}\!}
      \big( \braket{1^a 3} [2^b|P_{12}\ket{3}
          + \braket{2^b 3} [1^a|P_{12}\ket{3} \big) & \\
    - \sum_{k=4}^{n-1}
      \frac{ m \braket{k|k\!+\!1} \braket{3|1|P_{3 \dots k}|3}
             \bra{3}P_{3 \dots k}|
             \prod_{j=k}^{n-2}\!\big\{\!\!\not{\!P}_{13 \dots j}\!\not{\!p}_{j+1}
                                                + D_{13 \dots j} \big\}|n] }
           { s_{3 \dots k} \prod_{j=k}^{n-1} D_{13\dots j}\,
             \braket{3|1|P_{3 \dots k}|k} \braket{3|1|P_{3 \dots k}|k\!+\!1} } & \\ \times \big( \braket{1^a 2^b}
                   \braket{3|1|P_{3 \dots k}|3}
                 + \braket{1^a 3} \braket{2^b 3} s_{3 \dots k} \big) & \bigg]
\label{QQggnOM}
\eeal
involves a linearly growing number of terms.
However, all but one term contain the quark mass,
so only the first term survives the massless limit.
It gives rise to $2 \times 2 = 4$ helicity amplitudes,
two out of which vanish and the other two constitute
the quark counterparts of the Parke-Taylor formula~\eqref{MHV}.

\begin{figure}[t]
\flushright
\includegraphics[width = 0.9\textwidth,trim=0 0 10pt 0, clip=true]{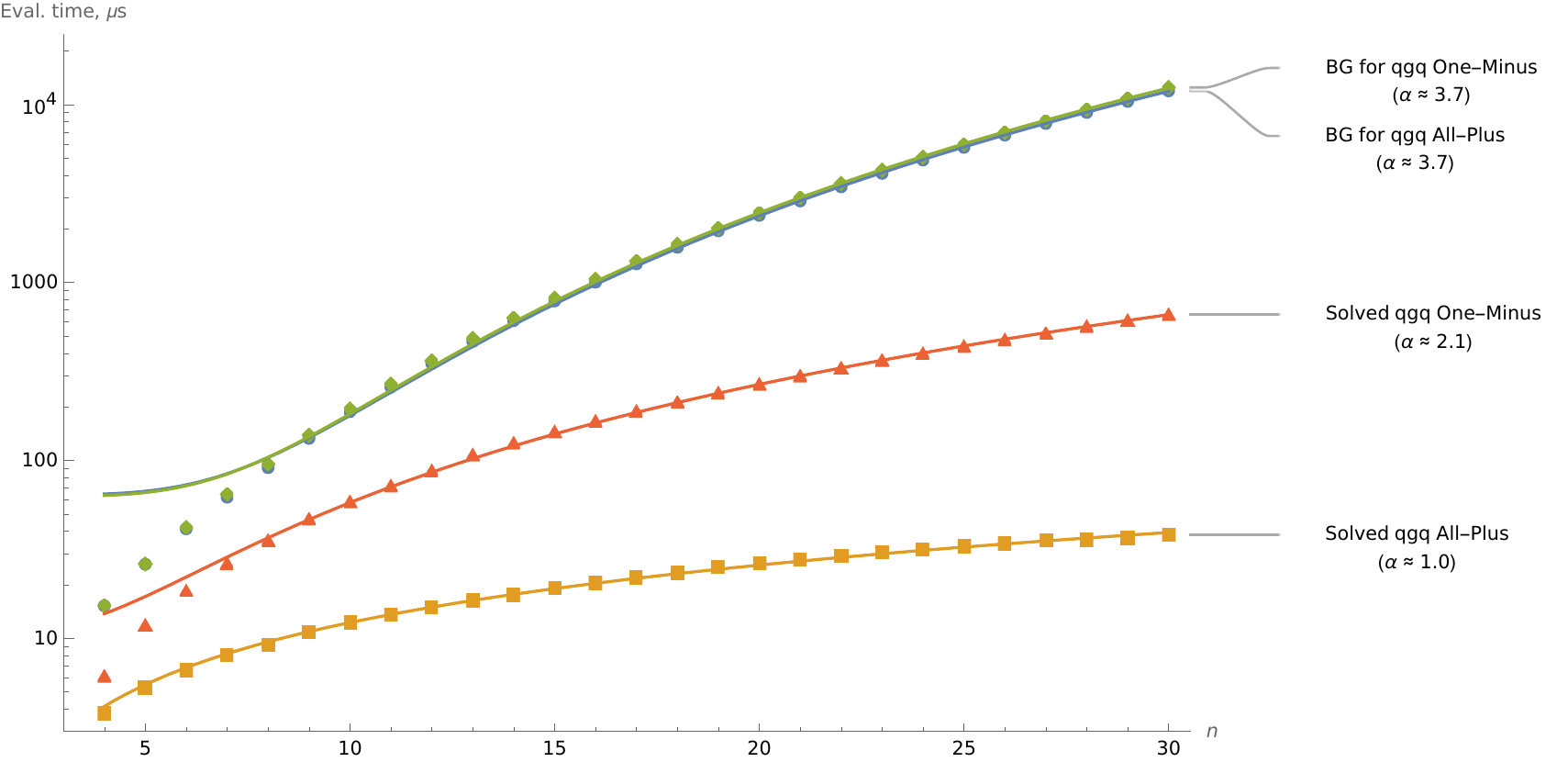}
\vspace{-3pt}
\caption{Evaluation times for $n$-point amplitudes involving two massive quarks. The time units are microseconds. The solid lines correspond to power-law fits of the form $a + b n_g^\alpha$, where $n_g=n-2$ is the number of gluons and the exponents $\alpha$ suggested by the data are shown in the legend above.}
\label{Plot2QLog}
\end{figure}

%%%%%%%%%%%%%%%%%%%%%%%%%%%%%%%%%%%%%%%%%%%%%%%%%%%%
\paragraph{Evaluation timings.}
In view of such a non-trivial analytic-complexity behavior
of our massive analytic results,
it is interesting how long it takes to evaluate them numerically.
\Figs{Plot2QLog}{Plot4QLog} show our timing results
for QCD amplitudes with two and four quarks, respectively.
The depicted timing data were obtained on a 2017 laptop
with a Intel Core i7-7500U processor and Ubuntu Linux 20.04.
Each amplitude was computed using its explicit formula
obtained by solving an on-shell BCFW recursion
and via an off-shell BG recursion.
During each measurement, the amplitudes were evaluated repeatedly
using double-precision floating-point numbers
on random kinematic phase-space points for approximately 10 seconds.
The time periods needed for each evaluation were averaged,
while their fluctuations resulted in error bars
that are too small to be visible in the figures behind the data-point markers.
As a sanity check, we have also verified that three other laptops
(equipped with macOS 11, 12 and Windows 11) produce qualitatively identical results.

\begin{figure}[t]
\flushright
\includegraphics[width = 0.9\textwidth,trim=0 0 10pt 0, clip=true]{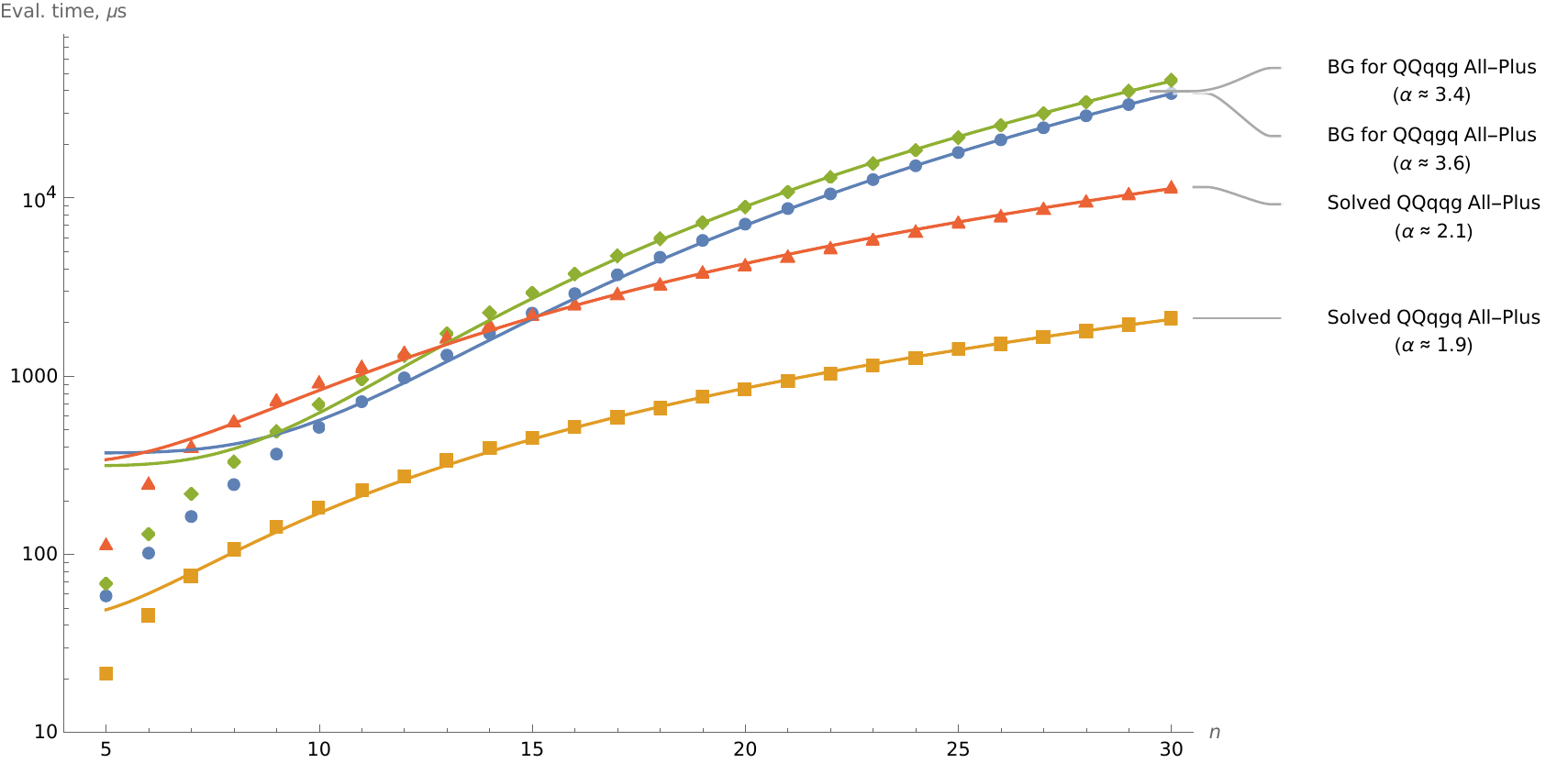}
\vspace{-3pt}
\caption{Evaluation times for $n$-point amplitudes involving four massive quarks. The time units are microseconds. The solid lines correspond to power-law fits of the form $a + b n_g^\alpha$, where $n_g=n-4$ is the number of gluons and the exponents $\alpha$ suggested by the data are shown in the legend above.}
\label{Plot4QLog}
\end{figure}

The two-quark amplitudes evaluated in \fig{Plot2QLog}
correspond to the ``all-plus'' formula~\eqref{QQggnAP}
involving only positive-helicity gluons
and to the ``one-minus'' formula~\eqref{QQggnOM}.
Both four-quark formulae~\eqref{eqAn1} and~\eqref{eqAn2}  in \fig{Plot4QLog}
correspond to all gluons having positive helicity
but in different positions with respect to the two quark flavors.
We have implemented these formulae using C++.
Both the analytic formulae and the BG recursion were implemented
for cross-checking purposes and were not fully optimized.
The BG recursion is implemented in a bottom-up approach,
used in~\rcite{Kanaki:2000ey},
and is similar to what is described in~\rcite{Badger:2012uz},
without the need for the cashing of off-shell currents
that was employed in~\rcite{Dinsdale:2006sq}.
The analytic formulae were implemented as given,
without low-level optimization.
As only the compiler optimization was used
(albeit with the ``aggressive'' flag \texttt{-Ofast}),
it is likely that some additional gain of perhaps up to 50\% in performance
is achievable for both approaches.
However, we do not expect this to affect neither the large-$n$ scaling
of the computational complexity
nor the comparison between algorithms to a significant degree.

There is an important difference in the way the analytic formulae
and the BG recursion are implemented.
That is, although the latter uses the same massive spinor parametrization
as the former, all the external spin labels are fixed
whenever the BG recursion routine is called,
since it is natural for an off-shell method
to treat massive and massless particles on the same footing.
This is in contrast to the on-shell analytic formulae,
which fix the massless helicities
but incorporate all massive spin degrees of freedom in one go ---
and which were implemented accordingly.
The formulae were therefore evaluated once
for each kinematic phase-space point,
whereas the BG recursion was called 4 and 16 times per point
in the two- and four-quark cases, respectively.

Our observations based on the evaluation timing data are the following.
\begin{itemize}
\item Evaluating the one-term all-plus expression~\eqref{QQggnAP}
is much faster than the BG recursion for any number of particles,
as clearly seen from \fig{Plot2QLog}.
The observed dependence on the total number of particles $n$ is roughly linear,
which is consistent with the growing complexity of the numerator
and denominator in the formula.

\item The one-minus formula~\eqref{QQggnOM} is also faster
than the BG recursion, but not by as large a margin
as the all-plus expression, see \fig{Plot2QLog}.
The observed dependence on the number of particles is roughly quadratic,
with different machines giving the slope $\alpha$ to be in the 1.9--2.1 range.
This is consistent with the growing number of terms
in combination with their increasing complexity.

\item The behavior of our simplest four-quark formula~\eqref{eqAn1}
with respect to the corresponding four-quark BG recursion,
as seen in \fig{Plot4QLog},
is qualitatively similar to that of the two-quark one-minus amplitude.
Namely, the analytic formula is faster
and has a milder growth curve than the BG recursion.

\item The more complicated four-quark formula~\eqref{eqAn2}
starts by taking roughly the same time to evaluate as the BG recursion
but due to a milder growth curve takes a lead from 9 external gluons ($n=13$),
see \fig{Plot4QLog}. On other machines, we have observed that
this intersection point may move (as high as $n=22$) but continues to exist.

\item The BG timings have a very mild dependence
on the particle configuration,
as is perhaps best seen in the unified \fig{PlotAll}.
Indeed, the power-law slopes $\alpha$ for all four recursive amplitude evaluations are in the region 3.4--3.7.
In fact, once we accounted for the factor of four difference
in the number of evaluations in the two- and four-quark cases,
we could observe that the four-quark amplitudes
with gluon insertions between like-flavored quarks
(labeled in the figures as ``QQqgq'')
are measurably faster to evaluate for the BG recursion
than the two-quark amplitudes, for the same number of particles.
This is because replacing two gluons by a quark pair may reduce
the number of possible factorization channels in an amplitude
and hence accelerate the recursion.
In the ``QQqgq'' configuration, this indeed happens because
the color ordering entirely prevents the external gluons from coupling
to one of the quark lines.

\item On the contrary, the analytic formulae depend very strongly
on the external particle and helicity configuration, see \fig{PlotAll}.
In particular, replacing gluons with quarks increases the evaluation time
by a factor increasingly greater than four,
at least in the case of identical-helicity gluons.

\begin{figure}[t]
\flushright
\includegraphics[width = 1.0\textwidth,trim=0 0 17pt 0, clip=true]{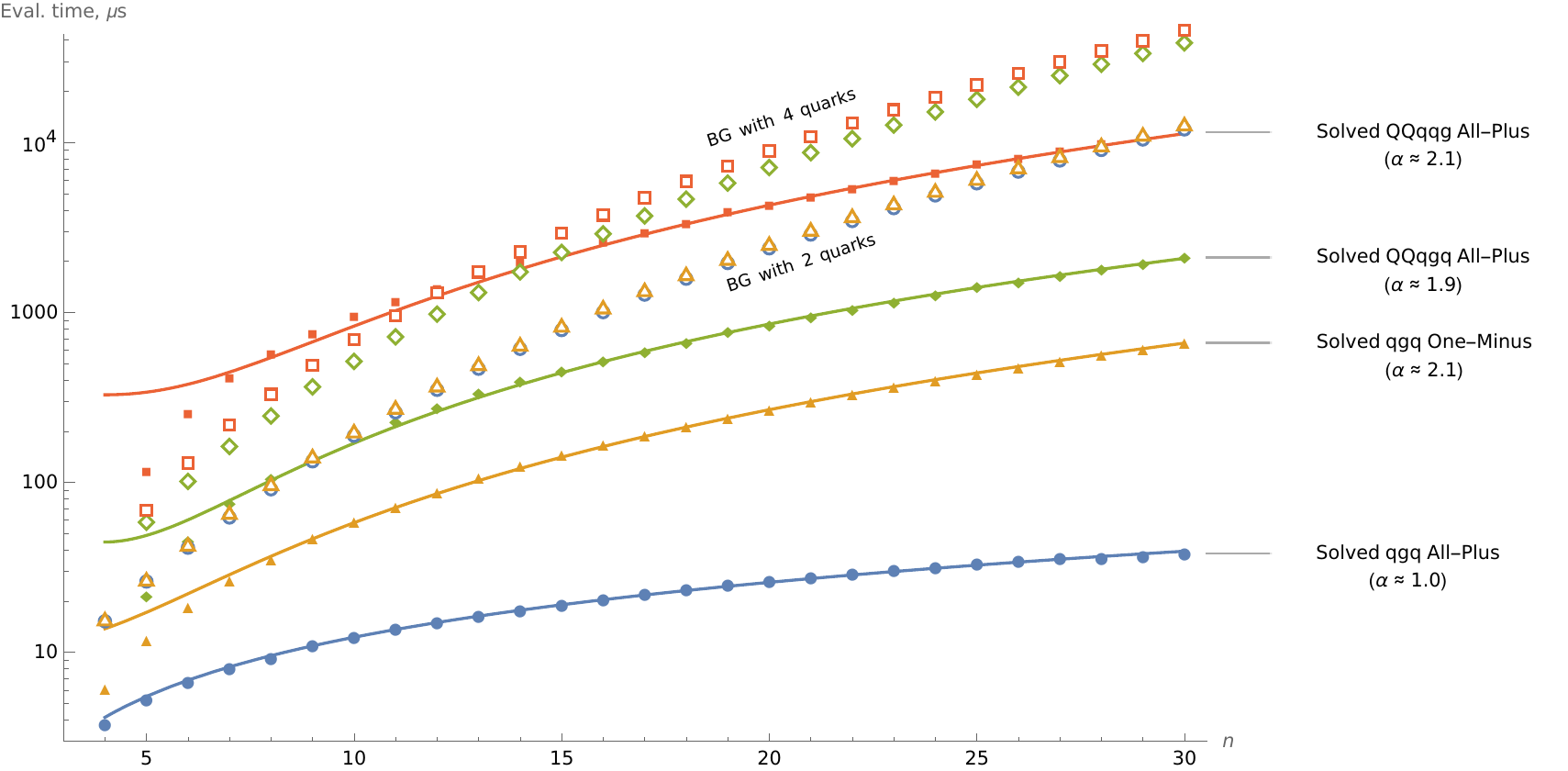}
\vspace{-3pt}
\caption{Evaluation times for all four $n$-point amplitudes involving two and four massive quarks. The time units are microseconds. The solid lines correspond to power-law fits and are not shown for the BG datapoints in view of their close proximity to each other.}
\label{PlotAll}
\end{figure}

\item However, the slopes of the four-quark analytics seem to be gentler
than expected.
Formula~\eqref{eqAn1} contains an increasing number of terms,
each of which grows in complexity with the number of particles~$n$,
so it might already seem slightly surprising that
its dependence on~$n$ appears to be milder than quadratic
in view of the observed $\alpha=1.9$.
This effect, however, is not very robust, 
measuring this slope on other machines produces results in the range 1.7--2.0,
which does include the expected behavior.
The more complicated amplitude~\eqref{eqAn2}, however,
even contains a double sum, which suggests a cubic dependence for large $n$,
as opposed to the observed $\alpha=2.1$.
This is definitely significantly tamer than 3
and is corroborated by our alternative datasets,
which place $\alpha$ in the range 1.7--2.1.
Our suspicion is that either our amplitude formula~\eqref{eqAn2}
responds particularly well to compiler optimization
or 30 particles is simply not enough
for these amplitude formulae to exhibit their true large-$n$ behavior .

\end{itemize}

%%%%%%%%%%%%%%%%%%%%%%%%%%%%%%%%%%%%%%%%%%%%%%%%%%%%
\section{Conclusions}
\label{sec:outro}
%%%%%%%%%%%%%%%%%%%%%%%%%%%%%%%%%%%%%%%%%%%%%%%%%%%%

In this paper, we have obtained new analytic all-multiplicity results
for gauge-theory amplitudes with massive matter.
First, we have substantiated the conjecture~\cite{Ochirov:2020lect}
for the amplitude~\eqref{MMggnAP} involving two massive spin-$s$ particles
and an arbitrary number of positive-helicity gluons
by showing that it is consistent with two different types of BCFW shifts.
This meshes particularly well with similar results \cite{Aoude:2020onz}
in heavy-particle effective theory~\cite{Damgaard:2019lfh,Aoude:2020onz}.
Then we have derived the closed formulae~\eqref{eqAn1} and~\eqref{eqAn2}
for two families of QCD amplitudes
involving two quark pairs and an arbitrary number of positive-helicity gluons.
The corresponding amplitudes with all negative-helicity gluons are
of course a collateral result.
Namely, they may be obtained simply by flipping the chirality
of all ${\rm SL}(2,\mathbb{C})$ spinors, \ie by exchanging
angle bra and ket spinors with square ket and bra spinors, respectively.

Finally, we have explored how much time our analytic formulae take
to be evaluated on a computer as opposed to
the widely used method of numerical off-shell recursion~\cite{Berends:1987me}.
We do not claim our implementations of either side of such a comparison
to have been sufficiently optimized to be production-ready,
but we still believe them to be adequate
for our purposes of qualitative comparison.
We find the large-multiplicity behavior of our amplitude formulae
to still be significantly milder than that of the BG recursion,
although the more involved of our two four-quark formulae
needs moderately high multiplicities (13--22 for our implementations)
to overtake the BG evaluation.
This means that solving the BCFW recursion in search of analytic formulae
in more complicated cases than considered here may hardly be motivated
by the need for numerically faster tree amplitudes.
Indeed, on-shell recursion relations with more terms
(or terms containing sums in themselves) are likely to produce analytic results
with three or more nested summations,
which will perform worse than the numerical off-shell recursion,
as previously observed in massless QCD \cite{Badger:2012uz}.\footnote{There is also the fact
that such an off-shell method as the BG recursion is much more flexible
with respect to different theories and particle configurations,
and so it is agnostic to the spin degrees of freedom,
on which the on-shell analytic results depend in a crucial way.
This flexibility argument admittedly often puts finding analytic solutions
at a disadvantage when it comes to production-ready codes.
}

We hope, however, that
the all-multiplicity formulae discussed in this paper
will provide new analytic data for the exploration of new structures of QCD.
For instance, it was the investigation of the geometric origin
of the spurious poles~\cite{Hodges:2009hk} in massless gauge-theory amplitudes
that has led to breakthroughs in our understanding
of the all-loop integrand structure of ${\cal N}=4$ supersymmetric
Yang-Mills theory~\cite{Arkani-Hamed:2009pfk,
Arkani-Hamed:2010zjl,Arkani-Hamed:2012zlh,Arkani-Hamed:2013jha},
such as its Grassmannian/Amplituhedron geometry.
In the massive-quark case at hand,
one could observe the presence of spurious poles already
in the one-minus two-quark amplitude formula~\rcite{Ochirov:2018uyq},
given in \eqn{QQggnOM}.
These poles appear in the form of composite spinor products
$\braket{3|P|Q|k}$, where $P$ and $Q$ are massive momenta,
which is very similar to what happens in massless NMHV amplitudes.
Now the four-quark formulae~\eqref{eqAn1} and~\eqref{eqAn2}
exhibit spurious poles of the new schematic form
$\big(D_P \bra{k}Q|c] + D_Q \bra{k}P|c]\big)$,
where $|c]$ is a composite spinor.
These poles seem less reminiscent of their massless counterparts,
and we wonder whether they could be understood
\eg in terms of twistor~\cite{Hodges:2009hk},
ambitwistor geometry~\cite{Mason:2013sva,Albonico:2021tbd}
and scattering equations~\cite{Cachazo:2013hca,Cachazo:2013iea,Dolan:2013isa,Dolan:2014ega,Naculich:2014naa,Naculich:2015coa,Naculich:2015zha,delaCruz:2015raa}.
We hope that finding such an understanding
of QCD amplitudes involving massive particles
will be facilitated by our new explicit analytic results.

\begin{acknowledgments}

We are very grateful to Alfredo Guevara, Jan Plefka, and Yael Shadmi for helpful discussions.
This research has received finding from the European Union's research and innovation programme Horizon 2020.
In particular, AO was funded under the Marie Sk{\l}odowska-Curie grant agreement 746138 and ERC grant \emph{PertQCD} (694712), and CS is funded under the Marie Sk{\l}odowska-Curie grant agreement 764850 ``SAGEX''.
AO's research is also funded by the STFC grant ST/T000864/1.

\end{acknowledgments}

\appendix
%%%%%%%%%%%%%%%%%%%%%%%%%%%%%%%%%%%%%%%%%%%%%%%%%%%%
\section{Spinor parametrizations}
\label{app:Parametrizations}
%%%%%%%%%%%%%%%%%%%%%%%%%%%%%%%%%%%%%%%%%%%%%%%%%%%%

Here we provide parametrizations for the spinor-helicity valuables
that may be consistently used for numerical evaluation of complex kinematics.
The subtle differences from those in \rcite{Ochirov:2018uyq}
were introduced mostly to enforce the momentum reversal rule
\be
   \ket{{-p}} = -\ket{p} , \qquad \quad |{-p}] = |p] ,
\label{MomentumReversal}
\ee
which now holds true for any complex massive or massless momenta.
For massless spinors, we choose following the parametrizations:
\begin{align}\!\!
    \ket{p}_\alpha & = I(p_+)
    \begin{pmatrix} ^{-\o{p}_{\perp}}\!/\!_{\sqrt{p_+}} \\ \sqrt{p_+}
    \end{pmatrix} , & &
    [p|_{\dot{\alpha}} = \frac{1}{I(p_+)}
    \begin{pmatrix} ^{-p_{\perp}}\!/\!_{\sqrt{p_+}} \\ \sqrt{p_+}
    \end{pmatrix}, & &
    \text{if } p_+ \neq 0 ; \\\!\!
    \ket{p}_\alpha & = I(p_-)
    \begin{pmatrix} -\sqrt{p_-} \\ ^{p_{\perp}}\!/\!_{\sqrt{p_-}}
    \end{pmatrix} , & &
    [p|_{\dot{\alpha}} = \frac{1}{I(p_-)}
    \begin{pmatrix} -\sqrt{p_-} \\ ^{\o{p}_\perp}\!/\!_{\sqrt{p_-}}
    \end{pmatrix} , & &
    \text{if } p_+=0 \text{ but } p_- \neq 0 ; \\\!\!
    \ket{p}_\alpha & = \frac{I(p^1)}{\sqrt{2p^1}}
    \begin{pmatrix} -\o{p}_{\perp}\! \\ p_{\perp}\!\!\!
    \end{pmatrix} , & &
    [p|_{\dot{\alpha}} = \frac{1}{I(p^1) \sqrt{2p^1}}
    \begin{pmatrix} -p_{\perp}\! \\ \o{p}_{\perp}\!\!\!
    \end{pmatrix} , & &
    \text{if } p_+=p_-=0 \text{ but } p^1\!\neq 0 .
\end{align}
Here we have used the light-cone variables
$p_\perp = p^1 + i p^2$, $\o{p}_\perp = p^1 - i p^2$ and $p_\pm = p^0 \pm p^3$,
with $p^2$ being the second spatial component of the momentum. 
We have defined the prefactors $I(p)$ in front of the parentheses as
\begin{align}
    I(p) = i^{\theta\left(^{i\sqrt{-p}}\!/\!_{\sqrt{p}}\right)}
         = \begin{cases}
             1 \qquad &\text{if}~~\sqrt{-p} /\! \sqrt{p} = i\\
             i \qquad &\text{if}~~\sqrt{-p} /\! \sqrt{p} = -i
         \end{cases}
\end{align}
to counter the uncertainty of treating the square root.
At a glance, the exponents may seem reducible to $\theta(i\sqrt{-1})$,
however, this would spoil the behavior under momentum reversal.

Massive spinors can be generically parametrized as
\beal\!\!
   \ket{p^a}_\alpha & = I(p_+)
      \Bigg\{
      \sqrt{\frac{E\!+\!P}{2P}}
      \begin{pmatrix} ^{-\o{p}_{\perp}}\!/\!_{\sqrt{p_+}} \\ \sqrt{p_+}
      \end{pmatrix}_{\!\alpha}\!\!\!\otimes\!
      \begin{pmatrix} 0 \\ 1
      \end{pmatrix}^{\!a}\!
    + \sqrt{\frac{E\!-\!P}{2P}}
      \begin{pmatrix} \sqrt{p_+} \\ ^{p_{\perp}}\!/\!_{\sqrt{p_+}}
      \end{pmatrix}_{\!\alpha}\!\!\!\otimes\!
      \begin{pmatrix} 1 \\ 0
      \end{pmatrix}^{\!a}\!\Bigg\} , \\\!\!
   [p^a|_{\dot{\alpha}} & =
      \frac{1}{I(p_+)}
      \Bigg\{
      \sqrt{\frac{E\!+\!P}{2P}}
      \begin{pmatrix} ^{-p_{\perp}}\!/\!_{\sqrt{p_+}} \\ \sqrt{p_+}
      \end{pmatrix}_{\!\dot{\alpha}}\!\!\!\otimes\!
      \begin{pmatrix} 1 \\ 0
      \end{pmatrix}^{\!a}\!
    - \sqrt{\frac{E\!-\!P}{2P}}
      \begin{pmatrix} \sqrt{p_+} \\ ^{\o{p}_{\perp}}\!/\!_{\sqrt{p_+}}
      \end{pmatrix}_{\!\dot{\alpha}}\!\!\!\otimes\!
      \begin{pmatrix} 0 \\ 1
      \end{pmatrix}^{\!a}\! \Bigg\} .\!\!\!
\label{MassiveSpinorSolution1}
\eeal
The variables used in \eqn{MassiveSpinorSolution1} are $E=p^0$,
$P = \sgn(E) \sqrt{\vec{\:\!p}^2}$ and $p_\pm = P \pm p^3$, where
\be
   \sgn(E) = \left\{
   \begin{aligned}
   +1 ,~ \text{ if } {\rm Re}(E) > 0 ,~ \text{ or if }
        {\rm Re}(E) = 0 \text{ and } {\rm Im}(E) \geq 0 ; \\
   -1 ,~ \text{ if } {\rm Re}(E) < 0 ,~ \text{ or if }
        {\rm Re}(E) = 0 \text{ and } {\rm Im}(E) < 0 .
   \end{aligned}
   \right.
\label{SignFunction}
\ee
Moreover, the identification
$\det\!\big\{\ket{p^a}_\alpha\big\}
=\det\!\big\{[p^a|_{\dot{\alpha}}\big\}=m>0$
can be spoiled by the determinants being negative
due to the square-root uncertainty again.
In that case, we restore their positivity
by introducing an additional factor of $i$ into $\ket{p^a}$
and $-i$ into $[p^a|$.

The above parametrization is invalid if $p_+=0$,
in which case we use
\beal
   \ket{p^a}_\alpha & = I(p_-)
      \Bigg\{
      \sqrt{\frac{E\!+\!P}{2P}}
      \begin{pmatrix} -\sqrt{p_-} \\ ^{p_{\perp}}\!/\!_{\sqrt{p_-}}
      \end{pmatrix}_{\!\alpha}\!\!\!\otimes\!
      \begin{pmatrix} 0 \\ 1
      \end{pmatrix}^{\!a}\!
    + \sqrt{\frac{E\!-\!P}{2P}}
      \begin{pmatrix} ^{\o{p}_{\perp}}\!/\!_{\sqrt{p_-}} \\ \sqrt{p_-}
      \end{pmatrix}_{\!\alpha}\!\!\!\otimes\!
      \begin{pmatrix} 1 \\ 0
      \end{pmatrix}^{\!a}\!\Bigg\} , \\
   [p^a|_{\dot{\alpha}} & = \frac{1}{I(p_-)}
      \Bigg\{
      \sqrt{\frac{E\!+\!P}{2P}}
      \begin{pmatrix} -\sqrt{p_-} \\ ^{\o{p}_\perp}\!/\!_{\sqrt{p_-}}
      \end{pmatrix}_{\!\dot{\alpha}}\!\!\!\otimes\!
      \begin{pmatrix} 1 \\ 0
      \end{pmatrix}^{\!a}\!
    - \sqrt{\frac{E\!-\!P}{2P}}
      \begin{pmatrix} ^{p_\perp}\!/\!_{\sqrt{p_-}} \\ \sqrt{p_-}
      \end{pmatrix}_{\!\dot{\alpha}}\!\!\!\otimes\!
      \begin{pmatrix} 0 \\ 1
      \end{pmatrix}^{\!a}\!\Bigg\} .\!\!
\label{MassiveSpinorSolution2}
\eeal
As before, additional factors of $\pm i$ should be introduced
to ensure positivity of the determinants.
The above two parametrizations fail when $P=0$,
in which case we have $p_\pm=\pm p^3$.
If it is not zero, we parametrize the massive spinors as
\beal
   \ket{p^a}_\alpha & = I(E - p_+)
      \Bigg\{\!
      \sqrt{\frac{\o{p}_\perp\!}{p_\perp\!}}
      \begin{pmatrix} \sqrt{E-p_+} \\
                      ^{-p_\perp}\!/\!_{\sqrt{E-p_+}}
      \end{pmatrix}_{\!\alpha}\!\!\!\!\otimes\!
      \begin{pmatrix} 0 \\ 1
      \end{pmatrix}^{\!a}\!\!
    + \frac{E}{\sqrt{E-p_+}}
      \begin{pmatrix} 0 \\ \sqrt{\,^{p_{\perp}}\!/_{\o{p}_\perp}\!}
      \end{pmatrix}_{\!\alpha}\!\!\!\!\otimes\!
      \begin{pmatrix} 1 \\ 0
      \end{pmatrix}^{\!a}\!\Bigg\} , \\
   [p^a|_{\dot{\alpha}} & = \frac{1}{I(E - p_+)}
      \Bigg\{\!
      \sqrt{\frac{p_\perp\!}{\o{p}_\perp\!}}
      \begin{pmatrix} \sqrt{E-p_+} \\
                     ^{-\o{p}_\perp}\!/\!_{\sqrt{E-p_+}}
      \end{pmatrix}_{\!\dot{\alpha}}\!\!\!\!\otimes\!
      \begin{pmatrix} 1 \\ 0
      \end{pmatrix}^{\!a}\!\!
    - \frac{E}{\sqrt{E-p_+}}
      \begin{pmatrix} 0 \\ \sqrt{\;\!^{\o{p}_{\perp}}\!/\!_{p_\perp}\!}
      \end{pmatrix}_{\!\dot{\alpha}}\!\!\!\!\otimes\!
      \begin{pmatrix} 0 \\ 1
      \end{pmatrix}^{\!a}\!\Bigg\} .\!\!
\label{MassiveSpinorSolution3}
\eeal
In the case where $E=p_+$, another option one can adopt is
\beal
   \ket{p^a}_\alpha & = I(E - p_-)
      \Bigg\{\!
      \sqrt{\frac{p_\perp\!}{\o{p}_\perp\!}}
      \begin{pmatrix} ^{-\o{p}_\perp}\!/\!_{\sqrt{E-p_-}} \\
                      \sqrt{E-p_-}
      \end{pmatrix}_{\!\alpha}\!\!\!\!\otimes\!
      \begin{pmatrix} 0 \\ 1
      \end{pmatrix}^{\!a}\!\!
    + \frac{E}{\sqrt{E-p_-}}
      \begin{pmatrix} \sqrt{\;\!^{\o{p}_{\perp}}\!/\!_{p_\perp}\!} \\ 0
      \end{pmatrix}_{\!\alpha}\!\!\!\!\otimes\!
      \begin{pmatrix} 1 \\ 0
      \end{pmatrix}^{\!a}\!\Bigg\} , \\
   [p^a|_{\dot{\alpha}} & = \frac{1}{I(E - p_-)}
      \Bigg\{\!
      \sqrt{\frac{\o{p}_\perp\!}{p_\perp\!}}
      \begin{pmatrix} ^{-p_\perp}\!/\!_{\sqrt{E-p_-}} \\
                      \sqrt{E-p_-} \\
      \end{pmatrix}_{\!\dot{\alpha}}\!\!\!\!\otimes\!
      \begin{pmatrix} 1 \\ 0
      \end{pmatrix}^{\!a}\!\!
    - \frac{E}{\sqrt{E-p_-}}
      \begin{pmatrix} \sqrt{\,^{p_{\perp}}\!/_{\o{p}_\perp}\!} \\ 0
      \end{pmatrix}_{\!\dot{\alpha}}\!\!\!\!\otimes\!
      \begin{pmatrix} 0 \\ 1
      \end{pmatrix}^{\!a}\!\Bigg\} .\!\!\!
\label{MassiveSpinorSolution4}
\eeal
A singular kinematics that invalidates the above parametrizations is
$P=p_\pm=0$. This means that either $p_\bot=0$ or $\overline{p}_\bot=0$,
and the momentum reduces to $p^\mu = \{ p^0, p^1, \pm i p^1, 0 \} $.
Then we use the following parametrization
\beal
   \ket{p^a}_\alpha & = I(p^1)
      \Bigg\{
      \frac{1}{\sqrt{2p^1}}
      \begin{pmatrix} -\o{p}_{\perp}\!+p^0 \\ p_{\perp}\!-p^0\!\!
      \end{pmatrix}_{\!\alpha}\!\!\!\otimes\!
      \begin{pmatrix} 0 \\ 1
      \end{pmatrix}^{\!a}\!
    + \frac{1}{\sqrt{2p^1}}
      \begin{pmatrix} \mp p^0 \\ \pm p^0
      \end{pmatrix}_{\!\alpha}\!\!\!\otimes\!
      \begin{pmatrix} 1 \\ 0
      \end{pmatrix}^{\!a}\!\Bigg\} , \\
   [p^a|_{\dot{\alpha}} & = \frac{1}{I(p^1)}
      \Bigg\{
      \frac{1}{\sqrt{2p^1}}
      \begin{pmatrix} -p_{\perp}\!\mp p^0 \\ \o{p}_{\perp}\!\mp p^0\!\!
      \end{pmatrix}_{\!\dot{\alpha}}\!\!\!\otimes\!
      \begin{pmatrix} 1 \\ 0
      \end{pmatrix}^{\!a}\!
    - \frac{1}{\sqrt{2p^1}}
      \begin{pmatrix} 2p_{\perp}\!-p^0 \\ 2\o{p}_{\perp}\!-p^0
      \end{pmatrix}_{\!\dot{\alpha}}\!\!\!\otimes\!
      \begin{pmatrix} 0 \\ 1
      \end{pmatrix}^{\!a}\!\Bigg\} .\!\!\!
\label{MassiveSpinorSolution5}
\eeal
Finally, for a massive particle at rest,
the momentum is $p^\mu = \{p^0, 0,0,0\} $,
in which case we may adopt the parametrization
that aligns the spin vectors with the $z$-axis,
\beal
   \ket{p^a}_\alpha & = I(p^0)
      \Bigg\{
      \begin{pmatrix} 0 \\ \sqrt{p^0}
      \end{pmatrix}_{\!\alpha}\!\!\!\otimes\!
      \begin{pmatrix} 0 \\ 1
      \end{pmatrix}^{\!a}\!
    + \begin{pmatrix} \sqrt{p^0} \\ 0
      \end{pmatrix}_{\!\alpha}\!\!\!\otimes\!
      \begin{pmatrix} 1 \\ 0
      \end{pmatrix}^{\!a}\!\Bigg\} , \\
   [p^a|_{\dot{\alpha}} & = \frac{1}{I(p^0)}
      \Bigg\{
      \begin{pmatrix} 0 \\ \sqrt{p^0}
      \end{pmatrix}_{\!\dot{\alpha}}\!\!\!\otimes\!
      \begin{pmatrix} 1 \\ 0
      \end{pmatrix}^{\!a}\!
    - \begin{pmatrix} \sqrt{p^0} \\ 0
      \end{pmatrix}_{\!\dot{\alpha}}\!\!\!\otimes\!
      \begin{pmatrix} 0 \\ 1
      \end{pmatrix}^{\!a}\!\Bigg\} .
\label{MassiveSpinorSolution6}
\eeal

%%%%%%%%%%%%%%%%%%%%%%%%%%%%%%%%%%%%%%%%%%%%%%%%%%%%
\section{Seed amplitudes for recursion}
\label{app:BuildingBlocks}
%%%%%%%%%%%%%%%%%%%%%%%%%%%%%%%%%%%%%%%%%%%%%%%%%%%%

Here we provide all the basic ingredients sufficient
for the QCD computations in this paper.
The three-gluon $\overline{\text{MHV}}$ amplitude is given by
the Parke-Taylor formula
\be
   A(1^+,2^+,3^-)
    = -\frac{i \sbraket{12}^3}{\sbraket{23} \sbraket{31}}
    =\!\!\!\!\!\includegraphics[valign=c,scale=0.9]{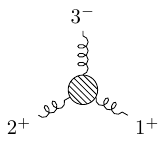}
     \!\!.
\label{MHVb3pt}
\ee
The amplitudes with two quarks and a gluon of either helicity are
\cite{Ochirov:2018uyq}
\be
   A(\u{1}^a,\o{2}^b,3^+)
    =\!\frac{i \braket{1^a 2^b} \sbraaket{3}{1}{q}}{m\braket{3\;\!q}}\!
    =\!\!\!\!\!\includegraphics[valign=c,scale=0.9]{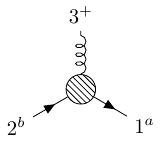}
   \quad
   A(\u{1}^a,\o{2}^b,3^-)
    =\!\frac{i \sbraket{1^a 2^b} \abrasket{3}{1}{q}}{m\sbraket{q\;\!3}}\!
    =\!\!\!\!\!\includegraphics[valign=c,scale=0.9]{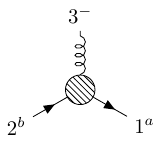}
     \!\!.
\label{eq3pt}
\ee
The last building block involves two pairs of distinctly flavored
massive quarks.
It is most easily computed from a single color-ordered Feynman diagram:
\begin{align}
\label{eq4Fermion}
   A(\u{1}^a,\o{2}^b,\u{3}^c,\o{4}^d) &
    = \includegraphics[valign=c,scale=0.9]{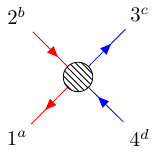}
    = \includegraphics[valign=c,scale=0.9]{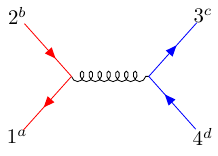} \\ &
    = \big({-\abra{1^a}}+\sbra{1^a}\big) \frac{-i \gamma_\mu}{\sqrt{2}}
      \big({-\aket{2^b}} + \sket{2^b}\big) \frac{-i}{s_{12}}
      \big({-\abra{3^a}}+\sbra{3^a}\big) \frac{-i \gamma^\mu}{\sqrt{2}}
      \big({-\aket{4^b}} + \sket{4^b}\big) \nn \\ &
    = \frac{i}{2 s_{12}}
      \Big( \abrasket{1^a}{\sigma_\mu}{2^b}
          + \sbraaket{1^a}{\bar{\sigma}_\mu}{2^b} \Big)
      \Big( \abrasket{3^c}{\sigma^\mu}{4^d}
          + \sbraaket{3^c}{\bar{\sigma}^\mu}{4^d} \Big) \nn \\ &
    = \frac{-i}{s_{12}}
      \Big( \braket{1^a 3^c}\sbraket{2^b 4^d}
          + \braket{1^a 4^d}\sbraket{2^b 3^c}
          + \sbraket{1^a 3^c}\braket{2^b 4^d}
          + \sbraket{1^a 4^d}\braket{2^b 3^c} \Big) . \nn
\end{align}
Here we have used the following construction of the Dirac spinors
from the Weyl spinors:
\be
   \bar{u}_p = \begin{pmatrix} -\abra{p} \\ \sbra{p} \end{pmatrix} , \qquad
   v_p = \begin{pmatrix} -\aket{p} \\ \sket{p} \end{pmatrix} , \qquad \quad
   \gamma^\mu = \begin{pmatrix} 0 & \sigma^{\mu} \\
               \bar{\sigma}^{\mu} & 0 \end{pmatrix} , \qquad
   \bar{\sigma}^{\mu,\dot{\alpha}\alpha} = \epsilon^{\alpha\beta}
      \epsilon^{\dot{\alpha}\dot{\beta}} \sigma^\mu_{\beta\dot{\beta}} .
\label{DiracSpinors}
\ee
We have also abused the notation by writing
$\bar{u}_p = -\abra{p} + \sbra{p}$ and $v_p = -\aket{p} + \sket{p}$,
which is allowed as long as one only contracts
indices of the same chirality.

%%%%%%%%%%%%%%%%%%%%%%%%%%%%%%%%%%%%%%%%%%%%%%%%%%%%
\section{Boundary behavior argument}
\label{app:BoundaryBehavior}
%%%%%%%%%%%%%%%%%%%%%%%%%%%%%%%%%%%%%%%%%%%%%%%%%%%%

Here we prove the vanishing boundary behavior of QCD amplitudes
under the massive quark-gluonic shift $[j^a,k\rangle$,
as given by \eqn{MassiveMasslessShift},
provided that the gluon $k$ has positive helicity.
The argument is similar the one in \rcite{Britto:2012qi}
for the massless shift.
In an arbitrary tree-level Feynman diagram,
we consider the unique path that the shift momentum $z r^\mu$ follows
from the antiquark $j$ to the gluon $k$:\footnote{Since our argument
targets the ${\cal O}(1)$ contributions,
we disregard the diagrams in which the path between the shifted
quark and gluon involves another fermion line:
such diagrams are ${\cal O}(1/z)$ by power counting.}
\be
\includegraphics[valign=c,scale=0.9]{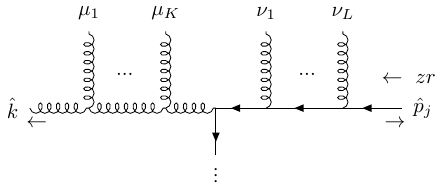} .
\label{ShiftedDiagram}
\ee
Here $K+L$ free Lorentz indices and the implicit fermion index
are understood to be contracted with other parts of the diagram,
which only depend on the sum $\hat{p}_j+\hat{p}_k=p_j+p_k$.

Let us examine how the leading term for large $z$ is constructed.
First of all, there are $K+L$ different propagator denominators
affected by the shift, which altogether provide
an ${\cal O}(1/z^{K+L})$ behavior.
The two shifted external wavefunctions compensate each other:
\be
   \hat{\varepsilon}_{k+}^\mu
    = \frac{\bra{q}\sigma^\mu|k]}{\sqrt{2} \braket{q\;\!\hat{k}}}
    = {\cal O}(1/z) , \qquad \quad
   \hat{v}_j^a = \begin{pmatrix} -\ket{j^a} \\ |\hat{j}^a] \end{pmatrix}
    = v_j^a - \frac{z}{m} |k] [k\;\!j^a]
    = {\cal O}(z) .
\ee
Moreover, on the left-hand side of the diagram, the term leading in $z$
is obtained from $K$ copies of the shift momentum $z r^\mu$
from each momentum-dependent gluonic vertex,
while the gluon propagator numerators are taken to be constant
in the Feynman gauge.
Finally, on the right-hand side of the diagram the vertices are constant,
whereas the highest power of $z$ is obtained by picking up
$L$ copies of $\;\not{\!\!\!\!\!zr}$ from each fermion propagator numerator.
Therefore, it seems like the diagram~\eqref{ShiftedDiagram}
should behave as ${\cal O}(1)$.

Before we can see that the leading ${\cal O}(1)$ term actually vanishes,
we note that it involves an odd Dirac-algebra element
$ \gamma^\nu \not{\!\!\!\!\!zr}\,\gamma^{\nu_1}\!\not{\!\!\!\!\!zr} \cdots
  \not{\!\!\!\!\!zr}\,\gamma^{\nu_L} $,
where the index $\nu$ is contracted with the $z$-dependent gluonic tree
on the left-hand side of the diagram.
This expression may be rewritten in terms of eight basic matrices
$\{\gamma^\mu,\gamma^{\mu} \gamma^5\}_{\mu=0,1,2,3}$ via the property
\be
   \gamma^\lambda \gamma^\mu \gamma^\nu
    = \eta^{\lambda \mu} \gamma^\nu - \eta^{\lambda \nu} \gamma^\mu
    + \eta^{\mu \nu} \gamma^\lambda
    - i \epsilon^{\lambda\mu\nu\rho} \gamma_\rho \gamma^5 .
\label{GammaReduction}
\ee
Thus the fermionic part of the ${\cal O}(1)$ contribution
must involve $\gamma^\mu|k]=\gamma^\mu\gamma^5|k]=\sigma^\mu|k]$.

Now we are ready to consider the faith of the Lorentz vector indices.
Altogether, there are $K+L$ free indices,
which may be distributed among $K+L$ copies of $z r^\mu$,
one polarization vector $\hat{\varepsilon}_{k+}^\mu$
and one Dirac-algebra structure $\gamma^\mu|k]$.
We see that at least some index contractions are required
to make this happen.
Since we have both the metric and Levi-Civita tensors in the mix,
the contractions allowed in the leading contribution are
\be\!\!
   \eta_{\mu\nu} V_1^\mu V_2^\nu, \quad
   \epsilon_{\lambda\mu\nu\rho} V_1^\mu V_2^\nu V_3^\rho ,
   \quad\!\text{where}\!\quad
   V_i^\mu \in \big\{ \underbrace{r^\mu\!\propto [k|\!\!\not{\!p}_j\sigma^\mu|k]}_{\text{up to }K+L\text{ copies}},~
     \hat{\varepsilon}_{k+}^\mu\!\!\propto \bra{q}\sigma^\mu|k],~
     \gamma^\mu|k]\,\big\} ,\!
\ee
both of which reduce the number of free indices by two.
However, all such combinations are exactly zero. Indeed,
$r^2 = r \cdot \hat{\varepsilon}_k^+ =\,\not{\!r}\,|k]
= \not{\!\hat{\varepsilon}_k^+}|k] = 0$, and the most involved check is for
\be
   \epsilon_{\lambda\mu\nu\rho}
   r^\mu \hat{\varepsilon}_{k+}^\nu \gamma^\rho |k]
    = - i\gamma^5 \big\{
        \gamma_\lambda\!\not{\!r}\!\not{\!\hat{\varepsilon}_k^+}
      - r_\lambda\!\not{\!\hat{\varepsilon}_k^+}
      + \hat{\varepsilon}_{k\lambda}^+\!\not{\!r}
      - (r \cdot \hat{\varepsilon}_k^+) \gamma_\lambda
        \big\} |k] = 0 ,
\ee
where we have reversed the property~\eqref{GammaReduction}
to remove the Levi-Civita contraction.
We may thus conclude that individual Feynman diagrams
behave as ${\cal O}(1/z)$ under the shift~\eqref{MassiveMasslessShift}.

%%%%%%%%%%%%%%%%%%%%%%%%%%%%%%%%%%%%%%%%%%%%%%%%%%%%
\section{Fermionic line reversal}
\label{app:FermionReversal}
%%%%%%%%%%%%%%%%%%%%%%%%%%%%%%%%%%%%%%%%%%%%%%%%%%%%

In this appendix we prove that reversing an arrow of a fermionic line
in a color-ordered amplitude results in an overall factor of $-1$.
In an arbitrary Feynman diagram, consider the part that directly
involves a fermionic line:
\be
\includegraphics[valign=c,scale=0.9]{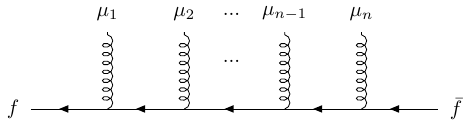} .
\label{FermionDiagram}
\ee
Here the Lorentz indices on the gluonic lines
(which for concreteness we put to the right of the fermion line)
should be contracted with other parts of the diagram,
each carrying a momentum flow $P_k$.
According to the color-ordered Feynman rules
and the Dirac spinors in \eqn{DiracSpinors},
the relevant part of the above diagram is then
\be
\big({-\abra{f}} + \sbra{f}\big)
\gamma^{\mu_1} (\not{\!\!P}_{f,1} + m)
\gamma^{\mu_2} (\not{\!\!P}_{f,2} + m) \dots
(\not{\!\!P}_{f,n-1} + m) \gamma^{\mu_n}
\big({-\aket{\bar{f}}} + \sket{\bar{f}}\big) ,
\label{FermionDiagramExpr}
\ee
where $P_{f,k} = p_f + \sum_{j=1}^k P_j$.
We omit the factors of $i$ and the propagator denominators.

We now wish to transpose the diagram
by putting the spinors $\bar{f}$ in the front and $f$ in the end.
Note that the number of gamma matrices between these spinors
vary between $n-1$ and $2n-1$,
the latter being the only term surviving the massless limit.
First, let us for concreteness consider this term,
which has an odd number of gamma (and therefore sigma) matrices.
Then we can transpose it directly by using the properties
\be
   \bra{f}^\alpha O_{\alpha\dot{\beta}} |\bar{f}]^{\dot{\beta}}
    = [\bar{f}|_{\dot{\beta}} (O^\mathsf{T})^{\dot{\beta}\alpha}
      \ket{f}_\alpha , \qquad \quad
   [f|_{\dot{\alpha}} O^{\dot{\alpha}\beta} \ket{\bar{f}}_\beta
    = \bra{f}^\beta (O^\mathsf{T})_{\beta\dot{\alpha}}
      |\bar{f}]^{\dot{\alpha}} .
\ee
Here transposition is understood to switch between
the two kinds of sigma matrices,
\be
   (\sigma_\mu^\mathsf{T})^{\dot{\alpha}\beta}
    = \bar{\sigma}_\mu^{\dot{\alpha}\beta}
    = \epsilon^{\dot{\alpha}\dot{\gamma}} \epsilon^{\beta\delta}
      \sigma_{\mu,\delta\dot{\gamma}} ,
   \qquad \quad
   (\bar{\sigma}_\mu^\mathsf{T})_{\alpha\dot{\beta}}
    = \sigma_{\mu,\alpha\dot{\beta}}
    = \epsilon_{\alpha\gamma} \epsilon_{\dot{\beta}\dot{\delta}}
      \bar{\sigma}_\mu^{\dot{\delta}\gamma} ,
\ee
as well as change the order of multiplication, \eg
\be
   \big( \sigma^\lambda \bar{\sigma}^\mu \sigma^\nu \big)^\mathsf{T}
    = \bar{\sigma}^\nu \sigma^\mu \bar{\sigma}^\lambda , \qquad \quad
   \big( \bar{\sigma}^\lambda \sigma^\mu \bar{\sigma}^\nu \big)^\mathsf{T}
    = \sigma^\nu \bar{\sigma}^\mu \sigma^\lambda , \qquad \quad \text{\etc.}
\ee
So it is clear that the massless term of \eqn{FermionDiagramExpr}
can be rewritten as
\be
\big({-\abra{\bar{f}}} + \sbra{\bar{f}}\big)
\gamma^{\mu_n} \not{\!\!P}_{f,n-1}
\gamma^{\mu_{n-1}} \not{\!\!P}_{f,n-2} \dots
\not{\!\!P}_{f,1} \gamma^{\mu_1}
\big({-\aket{f}} + \sket{f}\big) .
\ee

For an even number of gamma matrices,
the internal Weyl index structure is different,
and an additional sign appears due to internal contractions of the type
$\epsilon_{\alpha\gamma} \epsilon^{\beta\gamma} = -\delta_\alpha^\beta$:
\be
   \bra{f}^\alpha O_\alpha^{~\,\beta} \ket{\bar{f}}_\beta
    =-\bra{\bar{f}}^\beta (O^\mathsf{T})_\beta^{~\,\alpha} \ket{f}_\alpha ,
   \qquad \quad
   [f|_{\dot{\alpha}} O^{\dot{\alpha}}_{~\,\dot{\beta}}
      |\bar{f}]^{\dot{\beta}}
    =-[\bar{f}|_{\dot{\beta}} (O^\mathsf{T})^{\dot{\beta}}_{~\,\dot{\alpha}}
      |f]^{\dot{\alpha}} .
\ee
Note that such combinations of gamma matrices occur
in the Feynman diagram along with an odd power of mass~$m$,
whereas the odd combinations are multiplied by an even power of~$m$.
Therefore, we may absorb all the minuses induced by transposition
into the mass terms of the propagators.
Hence we rewrite the diagram~\eqref{FermionDiagramExpr} as
\begin{align} &
\big({-\abra{\bar{f}}} + \sbra{\bar{f}}\big)
\gamma^{\mu_n} (\not{\!\!P}_{f,n-1} - m) \gamma^{\mu_{n-1}}\!\dots
(\not{\!\!P}_{f,1} - m) \gamma^{\mu_1}
\big({-\aket{f}} + \sket{f}\big) \\ &\!= (-1)^{2n-1}
\big({-\abra{\bar{f}}} + \sbra{\bar{f}}\big)
(-\gamma^{\mu_n}) ({-\!\not{\!\!P}_{f,n-1}}\!+ m)
(-\gamma^{\mu_{n-1}}) \dots ({-\!\not{\!\!P}_{f,1}}\!+ m) (-\gamma^{\mu_1})
\big({-\aket{f}} + \sket{f}\big) . \nn
\end{align}
Here in the second line we have written the propagator numerators
in the usual form, as well as have accounted
for the flipped color-ordered Feynman rules,
which amounted to pulling out $(-1)^{2n-1} = -1$ to the front.
In this way, we may conclude that we have
examined all the consequences of flipping a fermion line
and arrived at precisely minus the fermion-flipped version
of the diagram~\eqref{FermionDiagram}, as promised.
Finally, note that the same sign flip would occur even if the gluonic lines
in the diagram were arranged differently.

%%%%%%%%%%%%%%%%%%%%%%%%%%%%%%%%%%%%%%%%%%%%%%%%%%%%
\section{BCFW recursive proofs}
\label{sec:proofeqA2}

%%%%%%%%%%%%%%%%%%%%%%%%%%%%%%%%%%%%%%%%%%%%%%%%%%%%
\subsection{5-point amplitude with gluon between distinctly flavored quarks}
\label{sec:proofeqA52}
%%%%%%%%%%%%%%%%%%%%%%%%%%%%%%%%%%%%%%%%%%%%%%%%%%%%

Here we present the derivation of the five-point amplitude~\eqref{eqA52}.
The first contribution~$B_5$ in \eqn{eq:diaA52}
is very similar to the five-point amplitude \eqref{grA12354},
where the gluon $5$ is inserted between the quarks $3$ and $4$.
Using the building blocks~\eqref{eq3pt} and~\eqref{eq4Fermion}, we find
\beal \label{eq:B5}
   B_5 &
    = \frac{i \sbraaket{{5}} {1} {q} \abraket{1|{-}\hat{P}_B^e} }
           {M D_{15} \abraket{\hat{5}q} \hat{s}_{34} }
      \big( \abraket{\hat{P}_{B,e}|3} \sbraket{2 \hat{4}}
          + \abraket{\hat{P}_{B,e}|4} \sbraket{2 3}
          + \sbraket{\hat{P}_{B,e}|3} \abraket{2 4}
          + \sbraket{\hat{P}_{B,e}| \hat{4}}\abraket{2 3} \big) \\ &
    = \frac{i \sbraaket{{5}} {1} {q}}
           {M D_{15} \abraket{\hat{5}q} \hat{s}_{34} }
      \big( M \abraket{13}[2\hat{4}] + M \abraket{14}[23]
          - \abrasket{1}{\hat{P}_{15}}{3} \abraket{24}
          - \abrasket{1}{\hat{P}_{15}}{\hat{4}} \abraket{23} \big) ,
\eeal
where we have contracted the spinors corresponding to
the internal momentum $\hat{P}_B$ in the same way as in \eqn{grA12354compute}.
In the rest of this appendix,
such an elimination of the intermediate momentum will be performed
without further explanations.
The shifted kinematics are evaluated at the pole where
$\hat{D}_{15} = \bra{\hat{5}}1|5] = 0$:
\begin{equation}  \label{eq:B5_pole}
   \frac{z}{m} =\frac{D_{15}}{[5|1|4|5]} , \qquad \quad
   \ket{\hat{5}} = -\frac{|1|5] D_{45}}{[5|1|4|5]} . %\qquad \quad
%   |\hat{4}^d] = |4^d] - \frac{D_{15}}{[5|1|4|5]} |5] [54^d] .
\end{equation}
Other shifted quantities, such as $|\hat{4}^d]$, can also be obtained
from this value of~$z$.
In particular, we have
\begin{equation} \label{eq:B5_3pt}
   \hat{s}_{34} = \frac{[5|1|P_{12}|3|P_{12}|5]}{[5|1|4|5]} , \qquad \quad
   \frac{[5|1\ket{q}}{\braket{\hat{5}q}} = \frac{[5|1|4|5]}{D_{45}} ,
\end{equation}
where the latter identity is true for any choice of $\ket{q}$
due to gauge invariance but is perhaps most obvious for $\ket{q}=|4|5]$.
Similarly, we will be fixing the reference spinors to our convenience
without further comments in the rest of the appendix.
Plugging \eqns{eq:B5_pole}{eq:B5_3pt} into \eqn{eq:B5},
we can show that the residue $B_5$ coincides with
the part of \eqref{eqA52} which is inside the big square brackets.

The contributions $C_5$ and $E_5$ are evaluated at the same pole,
where $\hat{s}_{34} = 0$ and hence
\begin{equation}\!\!\!
   \frac{z}{m} = \frac{\!-s_{34}}{[5|3|4|5]},   \qquad \quad
   \ket{\hat{5}} = \frac{|4|5] s_{12} + |P_{12}|5]D_{45}}{[5|3|4|5]} , \qquad \quad
%   |\hat{4}^d] = |4^d] + \frac{s_{34}}{[5|3|4|5]} |5] [54^d] , \qquad
   \ket{\hat{P}_C}[\hat{P}_C| = |P_{12}|5] \frac{[5|4|P_{34}|}{[5|3|4|5]} .
\label{eq:C5_pole}
\end{equation}
The contribution~$C_5$ involves an amplitude
with two quarks and with two positive-helicity gluons,
which is a special case of \eqref{QQggnAP}.
Combining with the three-point amplitude in \eqn{eq3pt}, we find
\begin{equation}
   C_5
    = -\frac{ i M \braket{12} \bra{\hat{P}_C}3|q] [3\hat{4}] [5|\hat{P}_C] }
            { m\:\!s_{34} \hat{D}_{15} [q|\hat{P}_C] \braket{\hat{5}|\hat{P}_C} }
    = \frac{ i M \braket{12} [3\hat{4}] }
           { m\:\!s_{34} \hat{D}_{15} }
      \frac{[5|P_{12}|3|5]}{\bra{\hat{5}}P_{12}|5]}.
\label{eqA5C}
\end{equation}
The necessary shifted quantities here are
\begin{equation}
   [3^c \hat{4}^d] = -\frac{m}{[5|3|4|5]}
      \big( [3^c 5] \bra{4^d}P_{12}|5] + [4^d 5] \bra{3^c}P_{12}|5]
      \big) , \qquad \quad
   \hat{D}_{15} = \frac{[5|1|P_{12}|3|P_{12}|5]}{[5|3|4|5]} .
\label{eqA5Caux}
\end{equation}
We emphasize that the mass $m$ in the denominator of \eqn{eqA5C}
is canceled by the numerator~$[3^c \hat{4}^d]$,
so there is no pole in the massless limit.
Using that $\bra{\hat{5}}P_{12}|5] = -s_{12}$ on this pole,
one can now easily verify that $C_5$
corresponds to the first line of \eqn{eqA52}.

Lastly, $E_5$ needs the four-point amplitude with two opposite-helicity gluons,
which can be found \eg in \rcite{Ochirov:2018uyq}:
\begin{equation}
   A(\u{1}^a,\o{2}^b,6^-,5^+) = \frac{i\bra{6}1|5]}{s_{12} D_{15}}
      \big( [1^a 5] \braket{2^b 6} + \braket{1^a 6} [2^b 5] \big) .
\end{equation}
With this we can write the contribution $E_5$ as
\begin{equation}
   E_5 = \frac{ i \braket{34} \bra{q}3|\hat{P}_C|1|5] }
              { m\:\!s_{12} s_{34} \hat{D}_{15} \bra{q}P_{12}|5] }
   \big( [15] \bra{2}P_{12}|5] + [25] \bra{1}P_{12}|5] \big).
   \label{eqA5D}
\end{equation}
The rest of the calculation involves using \eqns{eq:C5_pole}{eqA5Caux}
and produces the terms proportional to $m$ in the second line of \eqn{eqA52}.
This concludes our derivation of the five-point amplitude $A(1^a,2^b,3^c,4^d,5^+)$.

%%%%%%%%%%%%%%%%%%%%%%%%%%%%%%%%%%%%%%%%%%%%%%%%%%%%
\subsection{$\boldsymbol{n}$-point amplitude with gluons between distinctly flavored quarks}
\label{sec:proofeqAn2}
%%%%%%%%%%%%%%%%%%%%%%%%%%%%%%%%%%%%%%%%%%%%%%%%%%%%

Here we provide an inductive proof of the closed formula~\eqref{eqAn2}
for the $n$-point amplitude $A(1^a, 2^b,3^c,4^d,5^+,\dots ,n^+)$.
To set up the induction, we consider the base case $n=5$,
where only the first contribution~\eqref{eqAn21} survives.
Showing that it coincides with the five-point answer~\eqref{eqA52}
amounts to plugging in $|d_5^5] = |5]$ and $|e_5^d] = |4^d]$,
reducing the denominator $\bra{5}P_{12}|3|P_{125}|d_4^5]$ to
$s_{34} \bra{5}4|d_4^5] = s_{34}$
and using a couple of Schouten identities.

For general $n$-point amplitudes,
we are going to compute every diagram in \eqref{eqnptdia2}.
We will show that $B_n$, $C_n$ and $E_n$ correspond to the $i=5$ terms
in the sums~\eqref{eqAn22}, \eqref{eqAn23} and~\eqref{eqAn24}, respectively.
The rest of terms will be recursively generated by $F_n$.

The computation of $B_n$ is similar to that of the five-point case.
The crucial new step is to recognize
\begin{equation}
   \prod_{j=5}^{n-2}\!
   \big\{\!\!\not{\!\!P}_{(-P_B)5 \dots j}\!\not{\!p}_{j+1}
        + D_{(-P_B)5 \dots j} \big\} |n] = |a_6^n]
\end{equation}
in the numerator of the all-plus two-quark amplitude \eqref{MMggnAP}.
Gluing the latter with the four-quark amplitude \eqref{eq4Fermion}, we obtain
\begin{align}
   B_n &
    =-\frac{ iM \abraket{{-}\hat{P}_B^e|1} [5|a_6^n]
             \big( \abraket{\hat{P}_{B,e} 3} \sbraket{2 \hat{4}}
                 + \abraket{\hat{P}_{B,e} 4} \sbraket{2 3}
                 + \sbraket{\hat{P}_{B,e} 3} \abraket{2 4}
                 + \sbraket{\hat{P}_{B,e} \hat{4}} \abraket{2 3} \big) }
           { \prod_{k=6}^{n-1} \abraket{k|k\!+\!1} \abraket{\hat{5} 6}
             \prod_{l=5}^{n-1}\!D_{2345\dots l} D_{234} \hat{s}_{34} } \nn \\ &
    = \frac{ iM \sbraket{5|a_6^n}
             \big( M \braket{13} \sbraket{2\hat{4}} + M \braket{14} \sbraket{23}
                 + \abrasket{1}{\hat{P}_{234}}{3} \braket{24}
                 + \abrasket{1}{\hat{P}_{234}}{\hat{4}} \braket{23} \big) }
           { \hat{s}_{34} \braket{\hat{5} 6} \prod_{k=6}^{n-1} \braket{k|k\!+\!1}
             \prod_{l=4}^{n-1}\!D_{2\dots l} } .
\label{eq:Bn2}
\end{align}
This residue is evaluated at the pole $\hat{D}_{234}=0$, which implies
\begin{equation}\label{poleBn}\!\!
      \frac{z_B}{m} = \frac{D_{234}} {[5|4|P_{23}|5]}, \qquad
      |\hat{4}^d] = |4^d] - \frac{|5][54^d] D_{234}}{[5|4|P_{23}|5]} , \qquad
      \ket{\hat{5}} = \frac{|P_{234}|5] D_{45} - |4|5] D_{2345}}{[5|4|P_{23}|5]} .
\end{equation}
In particular, we have
\begin{equation}
    \hat{s}_{34} = \frac{[5|P_{34}|2|3|P_{34}|5]} {[5|4|P_{23}|5]} .
\end{equation}
Moreover, we can rewrite the shifted part of the numerator of \eqn{eq:Bn2} as
\begin{align}
  & M \braket{13} \sbraket{2\hat{4}} + M \braket{14} \sbraket{23}
  + \abrasket{1}{\hat{P}_{234}}{3} \braket{24}
  + \abrasket{1}{\hat{P}_{234}}{\hat{4}} \braket{23}  \\ &
  = M \braket{13} \sbraket{2\hat{4}} + M \braket{14} \sbraket{23}
  + \abrasket{1}{P_{2\dots 5}}{3} \braket{24}
  - \abraket{1\hat{5}} [53] \abraket{24}
  + \abrasket{1}{P_{2\dots 5}}{\hat 4} \braket{23}
  - \abraket{1\hat{5}} [54] \abraket{23} , \nn
\end{align}
which makes it easier to plug the shifted spinors~\eqref{poleBn}.
In this way, we find
\begin{align}
   B_n = -\frac{ iM [a_6^n|5] [5|P_{23}|4|5] }
                 { \prod_{j=6}^{n-1} \braket{j|j+1} [5|P_{34}|2|3|P_{34}|5]
                   \prod_{l=5}^{n-1} D_{2\dots l}
                   \big( D_{45} \abrasket{6}{P_{234}}{5}
                       - D_{2\dots 5} \bra{6}4|5] \big) } \quad & \\ \times
   \bigg\{ \abrasket{1}{P_{23}}{5} \braket{23} \sbraket{45}
         - \abrasket{1}{4}{5} \sbraket{35} \braket{24}
         + M \braket{13} \sbraket{25} \sbraket{45}
         - \frac{[5|P_{23}|4|5]} {D_{234}}
           \Big[ \abrasket{1}{P_{2\dots 5}}{3} \braket{24} & \nn \\
               + \abrasket{1}{P_{2\dots 5}}{4} \braket{23}
               + \braket{15} \braket{24} \sbraket{35}
               + \braket{15} \braket{23} \sbraket{45}
               + M \braket{13} \sbraket{24}
               + M \braket{14} \sbraket{23} &
           \Big]
   \bigg\} . \nn
\end{align}
To see that this accounts for the $i=5$ term in the sum~\eqref{eqAn22},
one needs to recall that
$|d_5^5] = |5]$, $|e_5] = |4]$ and $\bra{5}4|d_4^5] = 1$,
as well as use the statement
\begin{equation}
    |4^d] D_{45} = -m|5] \braket{4^d 5} - |4\aket{5} [4^d 5] ,
\end{equation}
which follows from the Schouten identity and the two-dimensional version
of the Dirac equation $|4|4^d] = m \ket{4^d}$.

The next two contributions $C_n$, $E_n$ come from the same factorization limit
in which the two quark pairs are entirely separated.
Gluing one of the three-point amplitudes in \eqn{eq3pt}
and the $(n-1)$-point two-quark amplitude \eqref{MMggnAP}
with all gluon helicities positive, we obtain
\begin{equation} \label{eqCn1}
   C_n
    =-\frac{ iM \braket{12} [3\hat{4}] \bra{\hat{P}_C}3|5]
             [\hat{P}_C| \big\{ \!\not{\!\!\hat{P}}_{2P_C}\!\not{\!\hat{p}}_5
                              + \hat{D}_{2P_C} \big\} |a_6^n] }
           { m s_{34} [5|\hat{P}_C] \braket{\hat{P}_C|\hat{5}}
            \braket{\hat{5} 6} \prod_{k=6}^{n-1} \braket{k|k\!+\!1}
            D_{2\hat{P}_C} \prod_{l=5}^{n-1} D_{2\dots l} } ,
\end{equation}
where we have again recognized the auxiliary spinor $|a_6^n]$.
The shifted quantities are evaluated at the pole where $\hat{s}_{34}=0$, where
\begin{equation}\!\!
    \frac{z_C}{m} = \frac{s_{34}}{[5|4|3|5]} , \qquad
    |\hat{4}^d] = \frac{ m |5] \bra{4^d}P_{34}|5] - |3|P_{34}|5] [4^d 5] }
                     { [5|3|4|5] } , \qquad
    \ket{\hat{5}} = \frac{|P_{345}|3|P_{34}|5]}{[5|3|4|5]} .
\label{eq:poleCn}
\end{equation}
In particular, we can then easily derive
\begin{equation}
    [3\hat{4}] = \frac{m}{[5|3|4|5]}
      \big( [35]\abrasket{4}{P_{34}}{5} + [45]\abrasket{3}{P_{34}}{5} \big) ,
\end{equation}
which cancels the factor of $m$ in the denominator.
Moreover, instead of eliminating $\hat{P}_C$ in \eqref{eqCn1},
we may substitute its massless spinors according to
\begin{equation} \label{eq:PC}
    \ket{\hat{P}_C} [\hat{P}_C| = -\frac{|P_{34}|5] [5|P_{34}|3|}{[5|3|4|5]} .
\end{equation}
It then follows that
\begin{equation}
   \hat{D}_{2P_C} = \frac{[5|P_{34}|2|3|P_{34}|5]}{[5|3|4|5]} , \qquad \quad
   \frac{ \bra{\hat{P}_C}3|5]
          [\hat{P}_C| \big\{\!\!\not{\!p}_2\!\not{\!\hat{p}}_5
                           + \hat{D}_{2P_C} \big\} |a_6^n] }
           { [5|\hat{P}_C] \braket{\hat{P}_C|\hat{5}} }
     = -\frac{[5|P_{34}|3|2|P_{345}|a_6^n]}{s_{345}} .
\end{equation}
Substituting all these shifted quantities in \eqref{eqCn1}
by the expressions given above,
we obtain the following compact form for $C_n$:
\begin{equation}
   C_n
    = \frac{ iM \braket{12} [5|3|4|5] [a_6^n|P_{345}|2|3|P_{34}|5]
             \big( [35]\abrasket{4}{P_{34}}{5}
                 + [45]\abrasket{3}{P_{34}}{5} \big)}
           { s_{34} s_{345} \bra{6}P_{345}|3|P_{34}|5] [5|P_{34}|2|3|P_{34}|5]
             \prod_{k=6}^{n-1}\braket{k|k\!+\!1}
             \prod_{l=5}^{n-1}\!D_{2\dots l} } .
\label{eqCn}
\end{equation}
This corresponds to the $i=5$ term in the sum~\eqref{eqAn23},
in which $\bra{5} P_{34}|3|P_{34}|d_4^5]$ is identified with $s_{34}$,
as follows from the definition~\eqref{eqspinord2}.

The contribution $E_n$ comes from the same factorization channel as $C_n$,
but the intermediate gluon has opposite helicity.
This means that $E_n$ involves
the two-quark amplitude with one negative-helicity gluon
and the rest of the gluon helicities being positive.
Such an amplitude was computed in \rcite{Ochirov:2018uyq}
and is given by \eqn{QQggnOM}.
In it, the minus-helicity gluon is color-adjacent to an outgoing quark,
whereas in $E_n$ it must stand next to an outgoing antiquark.
We flip the overall sign to reverse the arrow of the fermionic line,
as described in \app{app:FermionReversal},
and combine it with the first three-point amplitude in \eqn{eq3pt}.
This gives
\begin{align}
   E_n =\,&
      \frac{\!-i \braket{34} [\hat{P}_C|3\ket{5} }
           { m s_{34} \braket{\hat{P}_C|5} \braket{\hat{P}_C|\hat{5}}
            \prod_{j=6}^{n-1} \braket{j|j\!+\!1}\!}
      \bigg\{
      \frac{ \braket{\hat{P}_C|2|1|\hat{P}_C}
             \big( \braket{1|\hat{P}_C} [2|P_{12}\ket{\hat{P}_C}
                 + \braket{2|\hat{P}_C} [1|P_{12}\ket{\hat{P}_C} \big) }
           { s_{12} \braket{\hat{5} 6} \braket{\hat{P}_C|2|P_{12}|n} } \nn \\ +\,&
      \frac{ M \braket{\hat{P}_C|2|P_{345}|\hat{P}_C}
               \bra{\hat{P}_C}P_{345}|a_6^n]
             \big( \braket{12} \braket{\hat{P}_C|2|P_{345}|\hat{P}_C}
                 - \braket{1\hat{P}_C} \braket{2\hat{P}_C} s_{345} \big) }
           { s_{345} \prod_{j=5}^{n-1} D_{2\dots j}
             \braket{\hat{P}_C|2|P_{345}|\hat{5}} \braket{\hat{P}_C|2|P_{345}|6} }
\label{eq:En} \\
    + \sum_{k=6}^{n-1} &
      \frac{ M \braket{k|k\!+\!1} \braket{\hat{P}_C|2|P_{3\dots k}|\hat{P}_C}
               \bra{\hat{P}_C}P_{3\dots k}|a_{k+1}^n]
             \big( \braket{12} \braket{\hat{P}_C|2|P_{3\dots k}|\hat{P}_C}
                 - \braket{1\hat{P}_C} \braket{2\hat{P}_C} s_{3\dots k} \big) }
           { s_{3\dots k} \prod_{j=k}^{n-1} D_{2\dots j} \braket{\hat{5} 6}
             \braket{\hat{P}_C|2|P_{3\dots k}|k}
             \braket{\hat{P}_C|2|P_{3\dots k}|k\!+\!1} } \bigg\} , \nn
\end{align}
where we have once more recognized the appearance
of the auxiliary spinors $|a_{k+1}^n]$.
Using \eqns{eq:poleCn}{eq:PC} to substitute the shifted spinors,
we rewrite this contribution as
\begin{align}
   E_n =\,&
      \frac{ -im \braket{34} }
              { s_{34} s_{345} \prod_{j=6}^{n-1} \braket{j|j\!+\!1}\!}
         \bigg\{
         \frac{ [5|P_{34}|2|1|P_{34}|5]
                \big( \bra{1}P_{34}|5] [2|P_{12}|P_{34}|5]
                 \!+\!\bra{2}P_{34}|5] [1|P_{12}|P_{34}|5] \big) }
              { s_{12} \bra{n}P_{12}|2|P_{34}|5]
                \bra{6}P_{345}|3|P_{34}|5] }  \nn \\ &
         \qquad \qquad \qquad~\:\qquad
       + \frac{ M s_{345} [5|2|P_{34}|5] [a_6^n|5]
                \big( \braket{12} [5|P_{34}|2|5]
                    + \bra{1}P_{34}|5] \bra{2}P_{34}|5] \big) }
              { \prod_{l=5}^{n-1} D_{2\dots l}
                \bra{6}P_{345}|2|P_{34}|5]
                [5|P_{34}|2|3|P_{34}|5] } \nn \\
       + & \sum_{k=6}^{n-1}
         \frac{ M \braket{k|k\!+\!1} [5|P_{34}|2|P_{3\dots k}|P_{34}|5]
                      [a_{k+1}^n|P_{3\dots k}|P_{34}|5] }
              { s_{3\dots k} \prod_{j=k}^{n-1} D_{2\dots j}
                \bra{k}P_{3\dots k}|2|P_{34}|5]
                \bra{k\!+\!1}P_{3\dots k}|2|P_{34}|5]
                \bra{6}P_{345}|3|P_{34}|5] } \\* &
         \qquad \qquad \qquad \qquad\:\!\!\times\!
         \Big( \braket{12} [5|P_{34}|2|P_{3\dots k}|P_{34}|5]
             + s_{3\dots k} \bra{1}P_{34}|5] \bra{2}P_{34}|5] \Big)
         \bigg\} . \nn
\end{align}
Recalling that $\bra{5} P_{34}|3|P_{34}|d_4^5]=s_{34}$,
we can match it to the $i=5$ term in the sum~\eqref{eqAn24}.

Finally, the last contribution $F_n$ involves
a lower-point amplitude of the same type that we are calculating:
\begin{equation} \label{eq:channelFn}
   F_n = A(1^a,2^b,3^c,\hat{4}^d, \hat{P}_F^+, 7^+,...,n^+) \frac{-i}{s_{56}}
         A(-\hat{P}_F, \hat{5}^+, 6^+) ,
\end{equation}
so we need the inductive hypothesis.
Note that the following steps are largely analogous to those
following \eqn{eqAn1C} in the proof of the ordered amplitude~\eqref{eqAn1}.
This is because its structure is very similar to \eqn{eq:channelFn}.
Now the shifted kinematics in this channel is determined by
$\hat{s}_{56}=0$, which implies
\begin{equation}
    \frac{z_F}{m} = \frac{\braket{65}}{\bra{6}4|5]}, \qquad \quad
    |\hat{4}^d] = |4^d] - \frac{|5]\braket{56}[4^d 5]}{\bra{6}4|5]},\qquad \quad
    \aket{\hat{5}} = \frac{\ket{6}D_{45}}{\bra{6}4|5]} , 
\end{equation}
and the internal momentum $\hat{P}_F$ can be decomposed
into the following massless spinors:
\begin{equation}
    \ket{\hat{P}_F} = \ket{6}, \qquad \quad
    [\hat{P}_F| = [6| + \frac{D_{45}[5|}{\bra{6}4|5]}
%    = \frac{[5|( \not{\!\!P}_{45}\!\not{\!p}_6 + D_{45})}{\bra{6}4|5]}
    = \frac{[d_6^5|}{\bra{6}4|5]} ,
\label{IntermediateSpinors}
\end{equation}
where we recognized the appearance of $[d_6^5|$, as defined in \eqn{dspinor2}.
Meanwhile, the three-gluon $\overline{\text{MHV}}$ amplitude~\eqref{MHVb3pt} multiplied by the propagator gives
\begin{equation}
    \frac{-i}{s_{56}} A(-\hat{P}_F, \hat{5}^+, 6^+)
      = \frac{\bra{6}4|5]}{D_{45} \braket{56}} .
\end{equation}
This combines nicely with the prefactor
of the shifted $(n-1)$-point amplitude~\eqref{eqAn2}:
\begin{equation}
    \frac{\bra{6}4|5]}{D_{45} \braket{56}}
    \times \frac{\!-i}{\braket{\hat{P}_F|7} \prod_{j=7}^{n-1} \braket{j|j+1}}
    = \frac{\!-i}{\prod_{j=5}^{n-1} \braket{j|j+1}} \frac{\bra{6}4|5]}{D_{45}} ,
\label{eqAn2cPrefactor}
\end{equation}
where we see the formation of the $n$-point prefactor.

Now we consider the effects of the recursion on the terms of
the amplitude formula~\eqref{eqAn2} in the curly brackets,
which we assume to hold at $(n-1)$ points by the induction hypothesis.
Almost all the momentum sums appearing in the formula involve
both $\hat{p}_4$ and $\hat{P}_F$ and so remain unshifted:
\begin{equation}
     P_{4\dots j} ~\to~ \hat{P}_{4 P_F 7\dots j} = P_{4\dots j} , \qquad \quad
     D_{4\dots j} ~\to~ \hat{D}_{4 P_F 7\dots j} = D_{4\dots j} ,
\label{MomentumSumRecursion}
\end{equation}
where the arrows indicate the transition from the general $n$-point expression~\eqref{eqAn2} to the specific shifted $(n-1)$-point amplitude
$A(1^a, 2^b, 3^c, \hat{4}^d, \hat{P}_F, 7^+,\dots ,n^+)$.
The same obviously applies to
$P_{3\dots j} \to \hat{P}_{34 P_F 7\dots i} = P_{3\dots j}$ and to
$P_{2\dots j} \to \hat{P}_{234 P_F 7\dots i} = P_{2\dots j}$.
As for the auxiliary spinors,
one can use \eqns{IntermediateSpinors}{MomentumSumRecursion} to verify
\begin{equation}
   [d_i^5| ~\to~ [\hat{d}_i^{P_F}|
    = \frac{[d_6^5|}{\bra{6}4|5]}
    \prod_{k=6}^{i-1}
    \big\{\!\!\not{\!\!P}_{45 \dots k}\!\not{\!p}_{k+1}
    + D_{45 \dots k} \big\} = \frac{[d_i^5|}{\bra{6}4|5]}.
\end{equation}
whereas $|a_{i+1}^n]$ simply becomes itself.
One can further observe that in the formula~\eqref{eqAn2}
the spinor $|e_i]$ always appears contracted with $|5]$.
This spinor product becomes
\begin{equation}
   [5|e_i^d] ~\to~ [\hat{P}_F|\hat{e}_i^d] = \frac{[5|e_i^d]}{\bra{6}4|5]} .
\end{equation}

We are now ready to remove the hats from the residue~\eqref{eq:channelFn}.
Let us consider the terms coming from the $(n-1)$-point version of
the sum~\eqref{eqAn23} as a sample computation.
Plugging in the shifted kinematics and removing the hats
using the replacement rules derived above, we obtain
\begin{align}
 & \frac{ M \braket{12} \braket{67}
          \big( [3|d_6^5] \bra{4}P_{3456}|d_6^5]
              + \bra{3}P_{3456}|d_6^5] [e_i|5] \big)
          [d_6^5|P_{3456}|3|d_6^5] }
        { \bra{6}4|5]^2 s_{3456} D_{23456} \prod_{l=7}^{n-1} D_{2\dots l}
          \bra{7}P_{3456}|3|P_{3456}|d_6^5] \hat{s}_{34} }
   \frac{ [a_7^n|P_{3456}|2|3|P_{3456}|d_6^5] }
        { [d_6^5|P_{3456}|2|3|P_{3456}|d_6^5] } \nn \\ +
 & \sum_{i=7}^{n-1}
   \frac{ M \braket{12} \braket{i|i\!+\!1}
          \big( [3|d_i^5] \bra{4}P_{3\dots i}|d_i^5]
             + \bra{3}P_{3\dots i}|d_i^5] [e_i|5] \big)
          [d_i^5|P_{3\dots i}|3|d_i^5] }
        { \bra{6}4|5] s_{3\dots i} \prod_{l=i}^{n-1} D_{2\dots l}
          \prod_{k=6}^{i-1} D_{4\dots k}
          \bra{i\!+\!1}P_{3\dots i}|3|P_{3\dots i}|d_i^5]
          \bra{i}P_{3\dots (i-1)}|3|P_{3\dots (i-1)}|d_{i-1}^5]}
    \nn \\ &~\;\,\times\!
    \frac{ [a_{i+1}^n|P_{3\dots i}|2|3|P_{3\dots i}|d_i^5] }
    { [d_i^5|P_{3\dots i}|2|3|P_{3\dots i}|d_i^5]  }
    \label{eqAn2cShifted}
\end{align}
Here we have taken care to isolate the first term in the summation,
in which the factor $\bra{6}P_{34}|3|P_{34}|d_4^5] = s_{34}$
has been shifted to $\hat{s}_{34}$.
Then we can notice that
\begin{align}
    \hat{s}_{34} = \frac{\bra{6}P_{345}|3|P_{345}|d_{5}^5]}{\bra{6}4|5]} ,
\end{align}
so this term can be written as the $i=6$ term in the remaining sum.
In comparison with the terms~\eqref{eqAn23} in the $n$-point formula,
we seem still to be off by a factor of $\bra{6}4|5]/D_{45}$,
but it is exactly canceled by the prefactor \eqref{eqAn2cPrefactor}.
We have thus verified that the inductive residue $F_n$
converts the sum~\eqref{eqAn23} into itself but lacking the $i=5$ term,
which we have already retrieved from $C_n$.

Similar inductive arguments can be seen to hold
for the contribution~\eqref{eqAn21},
as well as for the sums~\eqref{eqAn22} and \eqref{eqAn24},
which are complemented by $B_n$ and $E_n$, respectively.
This concludes the proof of the $n$-point formula~\eqref{eqAn2}.

%%%%%%%%%%%%%%%%%%%%%%%%%%%%%%%%%%%%%%%%%%%%%%%%%%%%
\bibliographystyle{JHEP}
\bibliography{references}

\end{document}